\documentclass[reprint,twocolumn]{revtex4-2}
\usepackage{amsfonts}
\usepackage{amsmath}
\usepackage{amssymb}
\usepackage{charter}
\usepackage{graphicx}
\usepackage{float}

\begin{document}

\title{Phase sensitivity for an SU(1,1) interferometer via multiphoton
subtraction at the output port}
\author{Tao Jiang$^{1}$}
\author{Zekun Zhao$^{1}$}
\author{Qingqian Kang$^{1,2}$}
\author{Teng Zhao$^{1}$}
\author{Nanrun Zhou$^{3}$}
\author{Cunjin Liu$^{1,*}$}
\author{Liyun Hu$^{1,}$}
\thanks{Corresponding authors: lcjwelldone@126.com, hlyun@jxnu.edu.cn}
\affiliation{$^{{\small 1}}$\textit{Center for Quantum Science and Technology, Jiangxi Normal University, Nanchang 330022, China}\\
$^{{\small 2}}$\textit{Department of Physics, College of Science and Technology, Jiangxi Normal University, Nanchang 330022, China}\\
$^{{\small 3}}$\textit{School of Electronic and Electrical Engineering, Shanghai University of Engineering Science, Shanghai 201620, China}}

\begin{abstract}
In the field of quantum precision measurement, enhancing phase sensitivity
is crucial for various applications, including quantum metrology and quantum
sensing technologies. We theoretically investigate the improvement in phase
sensitivity and quantum Fisher information achieved through multiphoton
subtraction operations at the output port of an SU(1,1) interferometer under
conditions of photon loss. We use vacuum and coherent states as the inputs
and detect the outputs by intensity detection. The results indicate that
internal photon losses within the SU(1,1) interferometer have a more
significant impact on the phase sensitivity compared to external photon
losses. Moreover, increasing the number of photon subtractions $m$
effectively enhances both the phase sensitivity and the quantum Fisher
information. Notably, even under conditions of severe photon loss, the
multiphoton subtraction operations can enable the phase sensitivity to
surpass the standard quantum limit, approaching both the Heisenberg limit
and the quantum Cram\'{e}r-Rao bound. This study provides a new theoretical
basis for enhancing the phase sensitivity in the SU(1,1) interferometer.

\textbf{PACS: }03.67.-a, 05.30.-d, 42.50,Dv, 03.65.Wj
\end{abstract}

\maketitle

\section{Introduction}

Quantum precision measurement is a significant branch of quantum metrology
that utilizes quantum properties, such as quantum entanglement to achieve
measurement accuracy that surpasses classical limits \cite{1,2,3,4,5,6,7,8}.
It provides unprecedented accuracy in estimating unknown physical
parameters, which reveals widespread applications and rapid advancements in
gravitational wave detection \cite{b1,b2,b3}, optical imaging \cite{m1,m2,m3}%
, biological measurements \cite{g1,g2}, and fundamental physical
experiments. However, a primary challenge is to attain an accuracy exceeding
the standard quantum limit (SQL) \cite{9,10}. To address this challenge,
researchers have been developing innovative techniques and schemes to
further enhance the measurement accuracy and sensitivity \cite{11,12,13,14}.

Generally, the process of quantum precision measurement can be divided into
three stages: the preparation of quantum states, the interaction between the
interferometers and the input states, and finally, the measurement stage
\cite{16}. Researchers have been focusing on optimizing these stages to
achieve higher measurement accuracy. In classical theory, the measurement
limit is known as the SQL, which is defined by $1/\sqrt{N}$, where $N$
represents the average number of photons contributing to the measurement
\cite{7,17,19}. To achieve the phase sensitivity that surpasses the SQL,
various non-classical states such as entangled states \cite{20}, NOON states
\cite{21}, and twin Fock states, have been proposed as the light sources in
the measurement \cite{22}. Numerous studies have examined the use of
non-classical states to enhance the phase sensitivity of interferometers.
For instance, Caves proposed that introducing a squeezed vacuum state as the
input for the Mach-Zehnder interferometer (MZI) can enhance phase
sensitivity, allowing measurement accuracy to exceed the SQL \cite{7}.
However, high-quality non-classical states are not only difficult to
generate but also highly susceptible to environmental noise. Consequently,
utilizing the quantum properties of non-classical states to improve
measurement accuracy in noisy environments has become a significant focus of
research \cite{23,24,25}. For example, Xu $et$ $al.$ proposed utilizing
photon-added squeezed vacuum state and coherent state as inputs for the
SU(1,1) interferometer. Their research indicated that this approach can
surpass the SQL for phase sensitivity even in the presence of photon losses
\cite{26}. Similarly, Gong $et$ $al.$ proposed a scheme that employs
photon-subtracted squeezed vacuum state and coherent state as inputs for the
SU(1,1) interferometer, revealing significantly improved phase sensitivity
due to enhanced internal correlations within the interferometer \cite{27}.
Thus, the choice of input states plays a crucial role in influencing
measurement accuracy.

On the other hand, in the interaction process between quantum states and
interferometers, common MZI and SU(1,1) interferometer are employed. A
typical MZI consists of two $50\colon 50$ beam splitters (BSs) \cite%
{24,28,29}, while a standard SU(1,1) interferometer is composed of two
optical parametric amplifiers (OPAs) \cite{12,13,14,29}. As nonlinear
optical devices, OPAs are capable of generating quantum entanglement, which
significantly enhances both the signal-to-noise ratio and the phase
sensitivity, thereby making them highly effective in precision measurement
applications \cite{5,12,13}. Notably, several studies have demonstrated that
combining an MZI with an SU(1,1) interferometer can yield excellent results
in terms of phase sensitivity \cite{30,31,32}. Furthermore, extensive
investigations have been conducted on the use of various configurations and
different types of phase shifters to enhance the phase sensitivity of
interferometers \cite{a1,a2,a3,a4,a5}. Given the advantages of the SU(1,1)
interferometer, it has garnered considerable attention in both theoretical
and experimental research \cite{33,34,35,36}.

Although the SU(1,1) interferometer has demonstrated significant advantages
in phase estimation, studies by Stuart S. Szigeti $et$ $al.$ indicate that
only particles coupled through pumping contribute to the phase sensitivity
in the SU(1,1) interferometer, and the number of particles involved is
relatively small \cite{36}. To address this issue, researchers have
increasingly focused on performing non-Gaussian operations at the output
port of SU(1,1) interferometers to optimize phase encoding processes.
Notably, Zhang $et$ $al.$ investigated the enhancement of phase sensitivity
through single-photon subtraction at the output port of the SU(1,1)
interferometer under ideal conditions \cite{31}. To examine the impact of
the single-photon subtraction at the output on quantum Fisher information
(QFI), they proposed a non-Gaussian equivalent probe state that incorporates
the effects of the photon subtraction into the input state. In their
approach, the equivalent probe state relies on the unknown phase $\phi $ to
be measured. However, preparing such an input state using prior photon
subtraction or addition schemes is not feasible, especially for an unknown
phase-related case. To resolve this issue, we propose an equivalent model in
which the effects of the photon subtraction do not need to be incorporated
into the input state, allowing it to remain independent of the unknown phase
$\phi $, resulting in more reliable measurements. Our approach differs from
that of Zhang $et$ $al.$ in several key aspects. First, our proposed
equivalent model does not necessitate modifications to the input states and
is entirely based on the standard SU(1,1) interferometer, making it more
comprehensible and practical. Furthermore, while Zhang $et$ $al.$ focused on
the single-photon subtraction under ideal conditions, we emphasize the
crucial role of the multi-photon subtraction in the presence of photon
losses.

Our paper is organized as follows: In Sec. II, we introduce our model. In
Sec. III, we explore the enhancement of phase sensitivity in the SU(1,1)
interferometer through the multiphoton subtraction operations, considering
both ideal conditions and photon loss scenarios. In Sec. IV, we calculate
the QFI in the presence of photon losses and compare the phase sensitivity
with theoretical limits. Finally, in Sec. V, we present our conclusions.

\section{Model}

Our model is based on the standard SU(1,1) interferometer which consists of
two OPAs and a linear phase shifter, as illustrated in Fig. 1. In this
model, the input states comprise a vacuum state $\left \vert 0\right \rangle
_{a}$ and a coherent state $\left \vert \beta \right \rangle _{b}$ (where $%
\beta =\left \vert \beta \right \vert e^{i\theta _{\beta }}$, with $\theta
_{\beta }$ representing the phase). The input state can be expressed as $%
\left \vert \psi \right \rangle _{in}=\left \vert 0\right \rangle _{a}\otimes
\left \vert \beta \right \rangle _{b}$. Mode $a$ propagates through the two
OPAs and the linear phase shifter. The multiphoton subtraction operations
are implemented at the output port, and the phase sensitivity is measured
using intensity detection. Mode $b$ neither undergoes the phase shifter nor
the multiphoton subtraction operations. We illustrate the operational
process of the interferometer using mode $a$ as an example. The OPA process
can be represented by the two-mode squeezing operator $S_{k}^{\dagger }(\xi
_{k})=\exp \left( \xi _{k}ab-\xi _{k}a^{\dagger }b^{\dagger }\right) ,$
where $\xi _{k}=g_{k}e^{i\theta _{k}}$ denotes the squeezing parameter, with
$g_{k}$ and $\theta _{k}$ representing the gain factor and phase of the OPA,
respectively, with $k=1,2$. The operators $a$ $(a^{\dagger })$ and $b$ $%
(b^{\dagger })$ correspond to the annihilation (creation) operators for the
two modes. The phase shifter is described by the unitary operator $U_{\phi
}=e^{i\phi a^{\dagger }a}$, where $\phi $ denotes the phase shift. Finally,
we perform intensity detection at the output port of mode $a$.

For simplicity, we consider the parameters under the balance condition. For
the two OPAs, the balance condition is defined as $\theta _{1}=0,$ $\theta
_{2}=\pi $, and $g_{1}=g_{2}=g$ \cite{a7}. For the input coherent state of
mode $b$ $\left( \left \vert \beta \right \rangle _{b}\right) $, the phase
shift balance condition is $\theta _{\beta }=0$.

\begin{figure}[tph]
\label{Fig1} \centering \includegraphics[width=1.00\columnwidth]{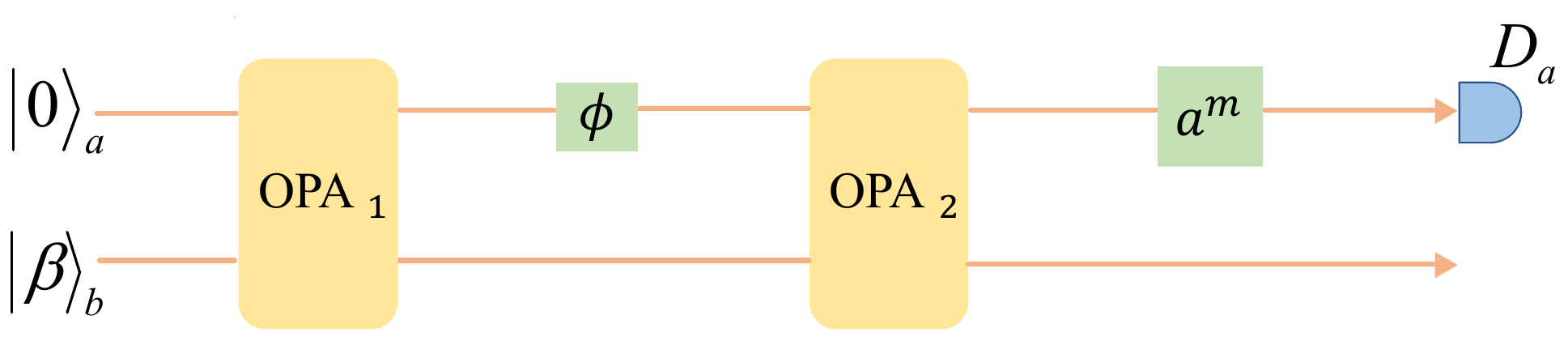}
\caption{{}Schematic diagram of our model based on SU(1,1) interferometer.
The two input ports of this interferometer are a vacuum state $\left \vert
0\right \rangle _{a}$ and a coherent state $\left \vert \protect \beta %
\right
\rangle _{b}$. OPA stands for optical parametric amplifier, $\protect%
\phi $ is the phase shift, and $D_{a}$ is the intensity detector. The
operator $a^{m}$ represents the subtraction of $m$ photons.}
\end{figure}

After undergoing a series of operations, the output state of mode $a$ before
detection can be expressed as

\begin{equation}
\left \vert \psi \right \rangle _{out}=N_{1}a^{m}S_{2}U_{\phi }S_{1}\left
\vert \psi \right \rangle _{in},  \label{0}
\end{equation}%
where $N_{1}$ is the normalization coefficient, which can be calculated as
\cite{26}

\begin{equation}
N_{1}=\left( G_{m}e^{A_{1}}\right) ^{-\frac{1}{2}},  \label{1}
\end{equation}%
and

\begin{equation}
G_{m}e^{\left( \cdot \right) }=\frac{\partial ^{2m}}{\partial t^{m}\partial
s^{m}}e^{\left( \cdot \right) }|_{t=s=0},  \label{b1}
\end{equation}%
\begin{equation}
A_{1}=st\left \vert w_{1}\right \vert ^{2}+\left( tw_{1}+sw_{1}^{\ast
}\right) \beta ,  \label{3}
\end{equation}%
as well as

\begin{equation}
w_{1}=\frac{1}{2}\sinh 2g\left( 1-e^{-i\phi }\right) .  \label{a2}
\end{equation}%
Here, $m$ represents the number of the photon subtraction operations, while $%
s$ and $t$ are differential variables. After differentiation, these
variables become $0$.

At the end of this section, we briefly discuss the feasibility of our
approach. In recent years, extensive research has focused on employing
non-Gaussian operations to enhance the phase sensitivity \cite%
{37,38,39,40,41,42}. Dakna $et$ $al.$ proposed that photon-added states,
such as photon-added thermal states, squeezed states, coherent states, and
displaced photon-number states, could be realized through conditional output
measurements at a BS \cite{44}. Furthermore, Namekata $et$ $al.$ have
experimentally demonstrated the photon subtraction operations using pulsed
squeezing at telecommunications wavelengths \cite{37}. Therefore, our
proposed schemes are feasible for implementing the photon subtraction
operations.

\section{Phase sensitivity based on intensity detection}

\subsection{Phase sensitivity in ideal case}

To discuss the phase sensitivity, three common detection methods are used,
i.e., intensity detection \cite{26,28,45}, homodyne detection \cite{46}, and
parity detection \cite{14,22,24}. Notably, intensity detection is both
experimentally feasible and straightforward for theoretical calculations.
Consequently, in this section, we employ intensity detection at the output
port of mode $a$ to estimate the phase sensitivity, which can be determined
using the error propagation formula \cite{29}

\begin{equation}
\Delta ^{2}\phi =\frac{\left \langle N^{2}\right \rangle -\left \langle
N\right \rangle ^{2}}{\left \vert \partial _{\phi }\left \langle N\right
\rangle \right \vert ^{2}}.  \label{a3}
\end{equation}

In this equation, $\left \langle N\right \rangle =\left. _{out}\left \langle
\psi \right \vert a^{\dagger }a\left \vert \psi \right \rangle _{out}\right.
$, where $N$ represents the number operator. According to Eq. (\ref{a3}), we
can obtain the phase sensitivity as follows (for detailed computation
process, see Appendix A) \cite{26}

\begin{eqnarray}
\Delta ^{2}\phi &=&\frac{N_{1}^{2}}{\left \vert \partial _{\phi }\left(
N_{1}^{2}G_{m+1}e^{A_{1}}\right) \right \vert ^{2}}  \notag \\
&&\times \lbrack G_{m+1}e^{A_{1}}+G_{m+2}e^{A_{1}}  \notag \\
&&-N_{1}^{2}(G_{m+1}e^{A_{1}})^{2}].  \label{a4}
\end{eqnarray}

\begin{figure}[tph]
\label{Fig2} \centering \includegraphics[width=0.83\columnwidth]{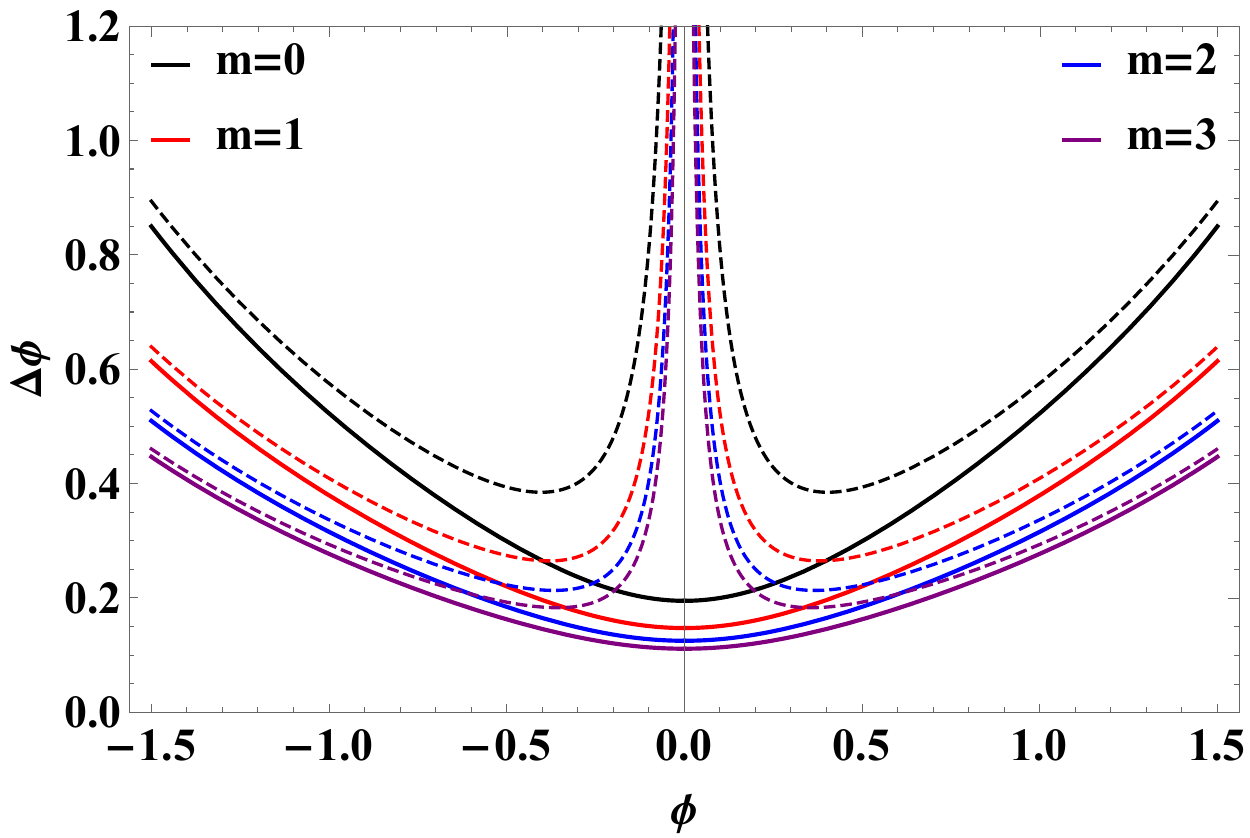}
\caption{{}The phase sensitivity based on intensity detection as a function
of $\protect \phi $ with $\protect \beta =1,$ $g=1$. The solid line and the
dashed line represent the detection of mode $a$ and mode $b$, respectively. $%
m$ is the number of the photon subtraction operations.}
\end{figure}

To investigate the influence of the multiphoton subtraction operations on
the phase sensitivity, we plot the phase sensitivity $\Delta \phi $ as a
function of $\phi $ as shown in Fig. 2. Here, the variable $m$ denotes the
number of the photon subtraction operations, with solid and dashed lines
representing detections at ports $a$ and $b$, respectively. From Fig. 2, we
can get the following conclusions: (i) when the parameters are chosen
identically, the detection results at port $a$ outperform those at port $b$;
(ii) at the same phase point, the phase sensitivity improves as $m$
increases; (iii) the intensity detection results for mode $a$ indicate that
the optimal phase sensitivity occurs near the phase point $\phi $ $=0$,
whereas this is not the case for mode $b$; (iv) when the phase shift is
large, the phase sensitivity detected at both ports is similar, but the
advantage is still kept by mode $a$.

\begin{figure}[tph]
\label{Fig3} \centering \includegraphics[width=0.83\columnwidth]{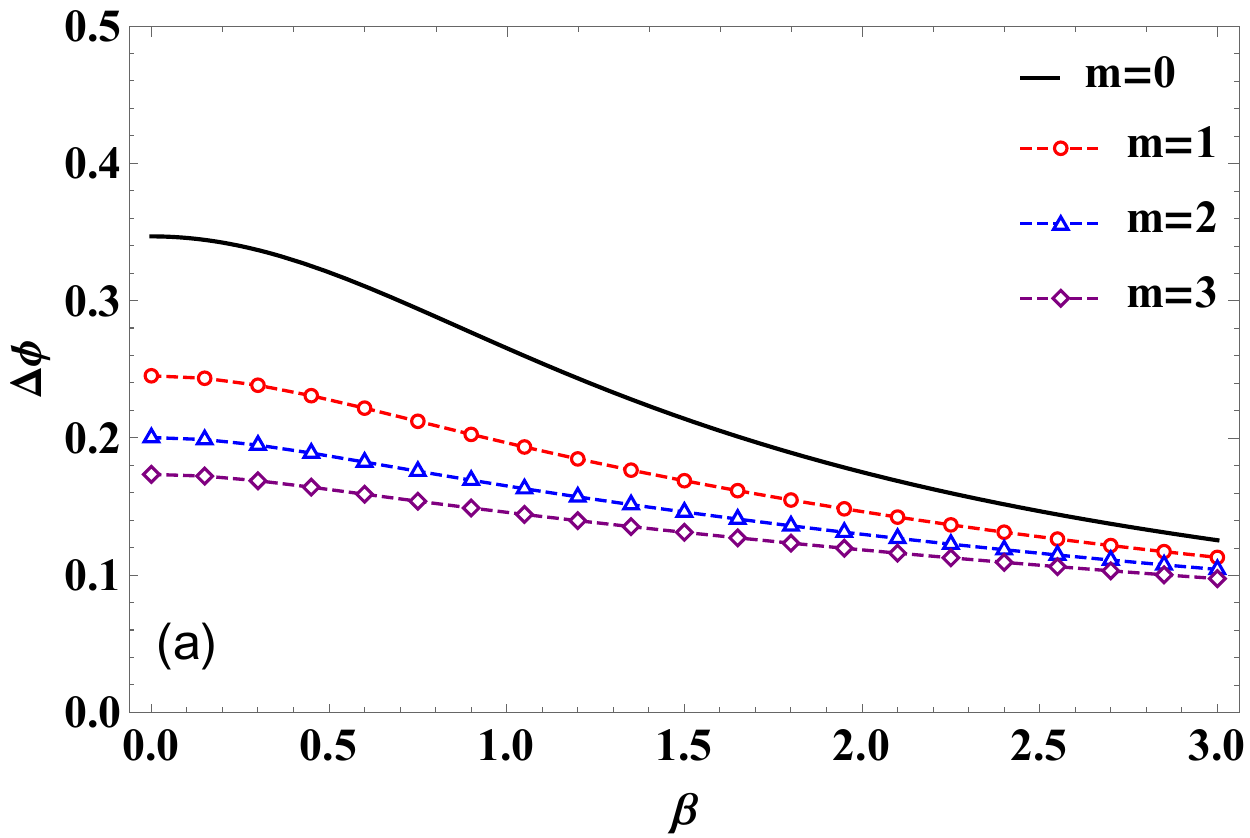} %
\centering \includegraphics[width=0.83\columnwidth]{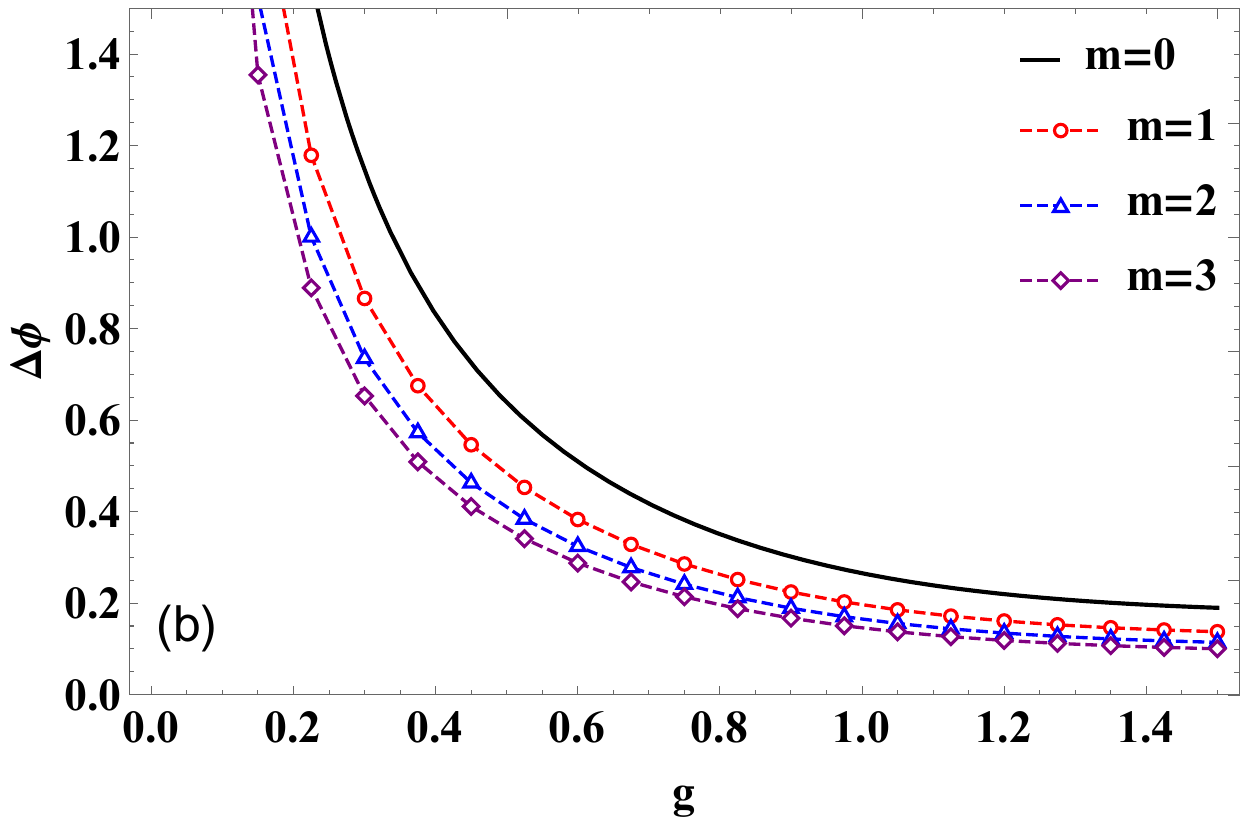}
\caption{{}The phase sensitivity based on intensity detection as a function
of (a) $\protect \beta $, $g=1$, $\protect \phi =0.4$; and (b) $g$, $\protect%
\beta =1$, $\protect \phi =0.4$.}
\end{figure}

Next, we explore the influence of other parameters on the phase sensitivity
of the interferometer. Given that the optimal phase point under conditions
of photon loss is not near $\phi $ $=0$, we will uniformly select $\phi $ $%
=0.4$ for the purposes of comparison in the following discussion. Fig. 3
illustrates the variations in the phase sensitivity $\Delta \phi $ with
respect to the coherent state amplitude $\beta $ of mode $b$ and the
amplification coefficient $g$ of the OPA. From Fig. 3, it is evident that:
(i) the phase sensitivity improves with $\beta $, attributing to the
increase in the number of photons entering the interferometer, which
enhances the energy of the light field, thereby enhancing the phase
sensitivity. Notably, this enhancement is more pronounced at lower values of
$\beta $, indicating that a relatively low energy input is sufficient to
achieve a high level of the phase sensitivity; (ii) at a constant $\beta $,
the phase sensitivity improves with $m$. Similar results can be obtained for
the case of gain factor $g$, see Fig. 3(b). This improvement can be
attributed to the OPA amplifying the signal power of the input light. As $g$
increases, the OPA enhances the amplification effect on the input signal,
resulting in a more robust output signal from the interferometer and thereby
enhancing the phase sensitivity.

As illustrated in Figs. 2 and 3, it is evident that the multiphoton
subtraction operations significantly enhance the phase sensitivity of the
SU(1,1) interferometer under ideal conditions. This improvement primarily
arises from the multiphoton subtraction operations, which result in an
increased photon count detected at the output port. Consequently, this
enhancement significantly improves both the signal strength and the
signal-to-noise ratio, thereby facilitating more accurate measurements of
phase changes.

\subsection{Phase sensitivity in the presence of photon losses}

\begin{figure}[tph]
\label{Fig4} \centering \includegraphics[width=1.00\columnwidth]{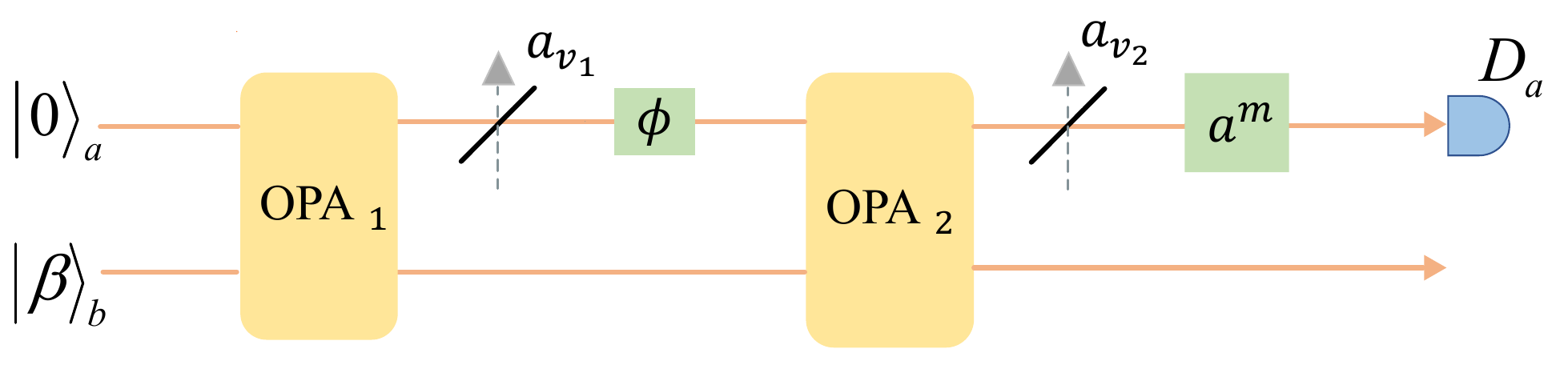}
\caption{{}Schematic diagram of the photon losses; adding fictitious BSs to
simulate photon losses inside and outside the SU(1,1) interferometer.}
\end{figure}

The preceding discussion is based on ideal conditions; however, in practical
applications, losses cannot be overlooked. Thus, we will analyze the effects
of both internal and external photon losses on the phase sensitivity
separately. We assume that photon losses occur between the first OPA and the
phase shifter, which we regard as internal losses. Conversely, photon losses
that occur between the second OPA and the multiphoton subtraction operations
are considered external losses, as shown in Fig. 4. To simulate these loss
processes, we introduce fictitious BSs, and their transformation relations
are expressed as

\begin{equation}
B_{k}^{\dagger }a_{k}B_{k}=\sqrt{T_{k}}a_{k}+\sqrt{1-T_{k}}a_{v_{k}},
\label{a5}
\end{equation}%
where $B_{k}$ is the BS operator, $a_{v_{k}}$ is the vacuum noise operator
corresponding to mode $a$, and $T_{k}$ denotes the transmittance of the BSs.
When $T_{k}=1$, it indicates no loss; when $T_{k}=0$, it indicates a hundred
percent loss, with $k=1,2$.

In the model that accounts for photon losses, the relationship between the
output state and the input state is given by

\begin{equation}
\left \vert \psi \right \rangle _{out}=N_{2}a^{m}B_{2}S_{2}U_{\phi
}B_{1}S_{1}\left \vert \psi \right \rangle _{in}\left \vert 0\right \rangle
_{a_{v_{1}}}\left \vert 0\right \rangle _{a_{v_{2}}},  \label{a6}
\end{equation}%
where $N_{2}$ is the normalization coefficient, and it can be expressed as

\begin{equation}
N_{2}=\left( G_{m}e^{A_{2}}\right) ^{-\frac{1}{2}}.  \label{a61}
\end{equation}

Based on Eqs. (\ref{a3}), (\ref{a5}) and (\ref{a6}), we can obtain the phase
sensitivity $\Delta ^{2}\phi _{L}$ under photon losses, i.e.,

\begin{eqnarray}
\Delta ^{2}\phi _{L} &=&\frac{N_{2}^{2}}{\left \vert \partial _{\phi }\left(
N_{2}^{2}G_{m+1}e^{A_{2}}\right) \right \vert ^{2}}  \notag \\
&&\times \lbrack G_{m+1}e^{A_{2}}+G_{m+2}e^{A_{2}}  \notag \\
&&-N_{2}^{2}(G_{m+1}e^{A_{2}})^{2}],  \label{a7}
\end{eqnarray}%
where%
\begin{equation}
A_{2}=st\left \vert w_{3}\right \vert ^{2}+\left( tw_{3}+sw_{3}^{\ast
}\right) \beta ,  \label{c1}
\end{equation}%
and%
\begin{equation}
w_{3}=\frac{1}{2}\sinh 2g\sqrt{T_{2}}(1-\sqrt{T_{1}}e^{-i\phi }).  \label{c2}
\end{equation}%
\begin{figure}[tph]
\label{Fig5} \centering \includegraphics[width=0.83\columnwidth]{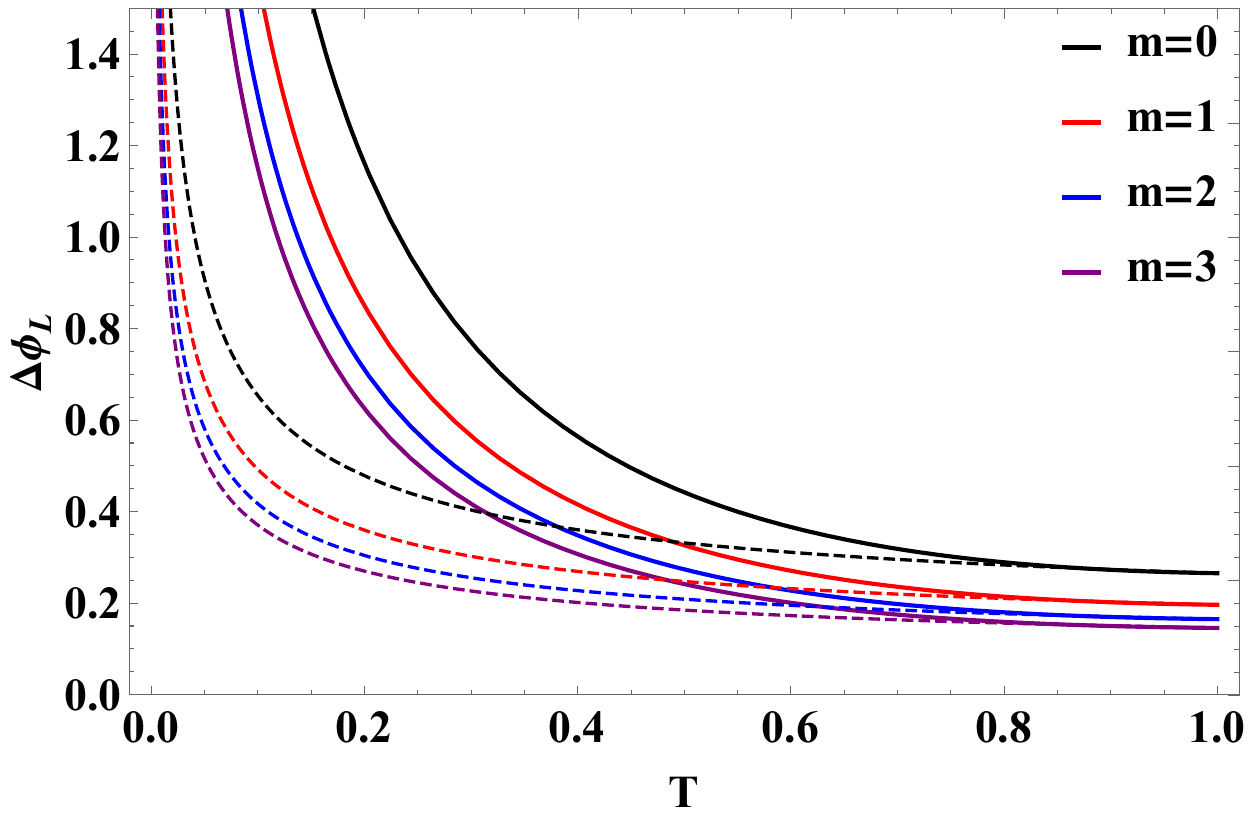}
\caption{{}The phase sensitivity based on intensity detection as a function
of $T$ with $\protect \beta =1$, $\protect \phi =0.4$, and $g=1$. The solid
and the dashed lines represent the conditions with internal and external
photon losses, respectively.}
\end{figure}

From Eqs. (\ref{a7})-(\ref{c2}), it is evident that when $T_{1}=T_{2}=1,$
the phase sensitivity with photon losses returns to that of the ideal case.
Next, we will explore the effects of internal $\left( T_{2}=1,T_{1}=T\right)
$ and external $\left( T_{1}=1,T_{2}=T\right) $ photon losses on the phase
sensitivity separately. Fig. 5 illustrates the variation of the phase
sensitivity $\Delta \phi _{L}$ as a function of $T$ in the presence of
photon losses. From Fig. 5, we can draw the following conclusions: (i) both
internal and external photon losses adversely affect the phase sensitivity;
however, internal photon losses have a more significant impact, particularly
at lower $T$. This can be attributed to the model, as photon losses inside
the interferometer are amplified by the second OPA, whereas external losses
are not, resulting in a more pronounced effect of internal losses on the
phase sensitivity; (ii) the photon subtraction operations can enhance the
phase sensitivity of the SU(1,1) interferometer in a noisy environment, with
the degree of improvement increasing with the number of photons subtracted.

From the above discussion, it is clear that internal photon losses within
the interferometer significantly affect the phase sensitivity in our model.
Consequently, we will focus on the case of internal photon losses in the
subsequent analysis.

\section{QFI and theoretical limits}

\subsection{QFI in ideal case}

The QFI is a crucial metric for assessing the maximum information contained
in an unknown phase shift $\phi $, within the realm of quantum precision
measurement. The quantum Cram\'{e}r-Rao bound (QCRB) associated with the QFI
represents the ultimate theoretical limit for phase measurement. This bound
is independent of the detection method employed, and for fixed input
resources, the QFI remains constant. In the ideal case, for a pure state as
the input, the QFI is given by

\begin{equation}
F=4\left[ \left \langle \psi _{\phi }^{^{\prime }}|\psi _{\phi }^{^{\prime
}}\right \rangle -\left \vert \left \langle \psi _{\phi }^{^{\prime }}|\psi
_{\phi }\right \rangle \right \vert ^{2}\right] ,  \label{a111}
\end{equation}%
and the QCRB is given by%
\begin{equation}
\Delta \phi _{QCRB}=\frac{1}{\sqrt{\nu F}},  \label{a8}
\end{equation}%
where $\nu $ represents the number of repeated experiments. For simplicity,
we set $\nu =1.$
\begin{figure}[tph]
\label{Fig6} \centering \includegraphics[width=1\columnwidth]{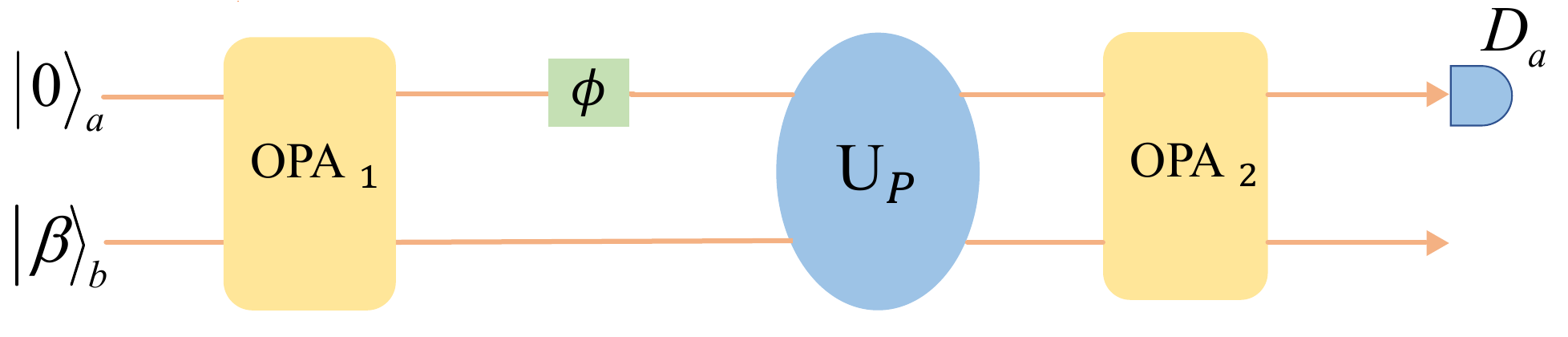}
\caption{{}Schematic diagram of the equivalent model of the SU(1,1)
interferometer, where $U_{P}$ represents a non-local operation, and $%
U_{P}=S_{2}^{\dagger }a^{m}S_{2}$.}
\end{figure}
\begin{figure}[tph]
\label{Fig7} \centering \includegraphics[width=0.83\columnwidth]{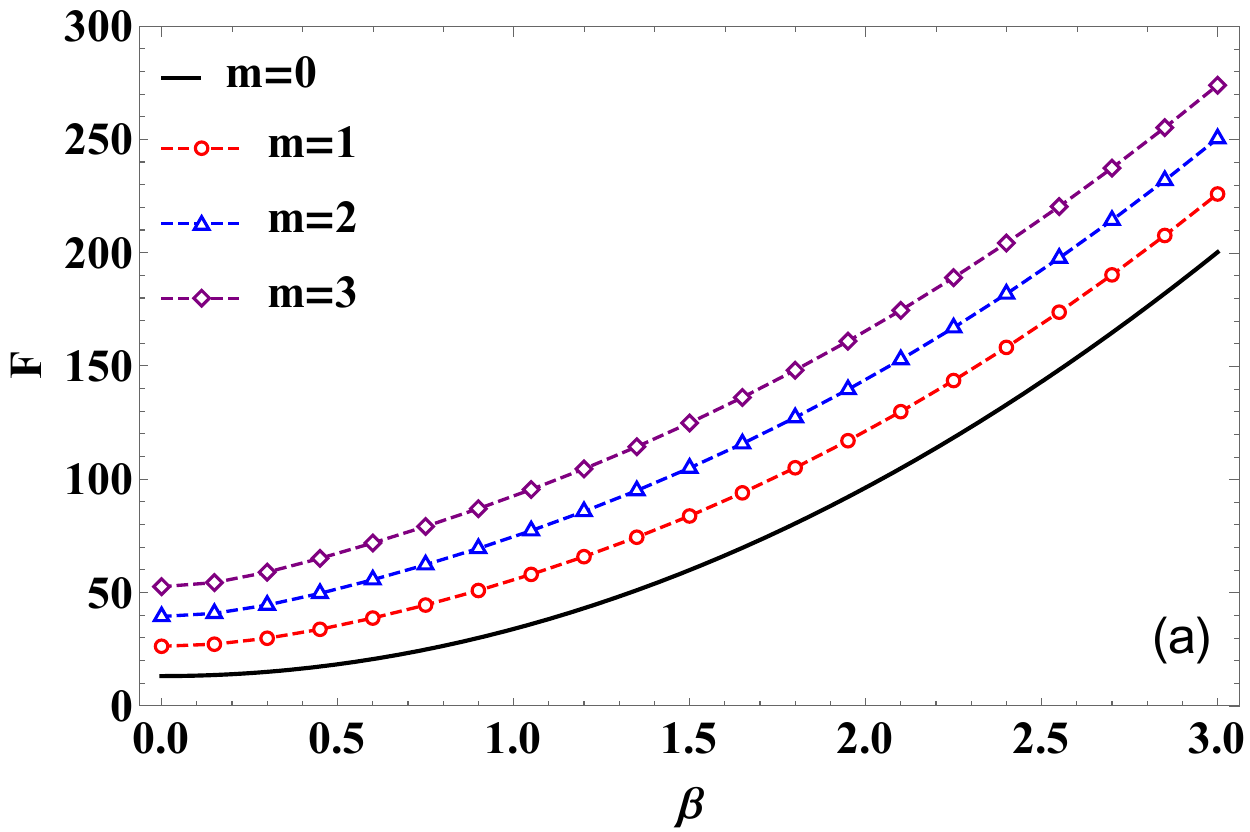} %
\centering \includegraphics[width=0.83\columnwidth]{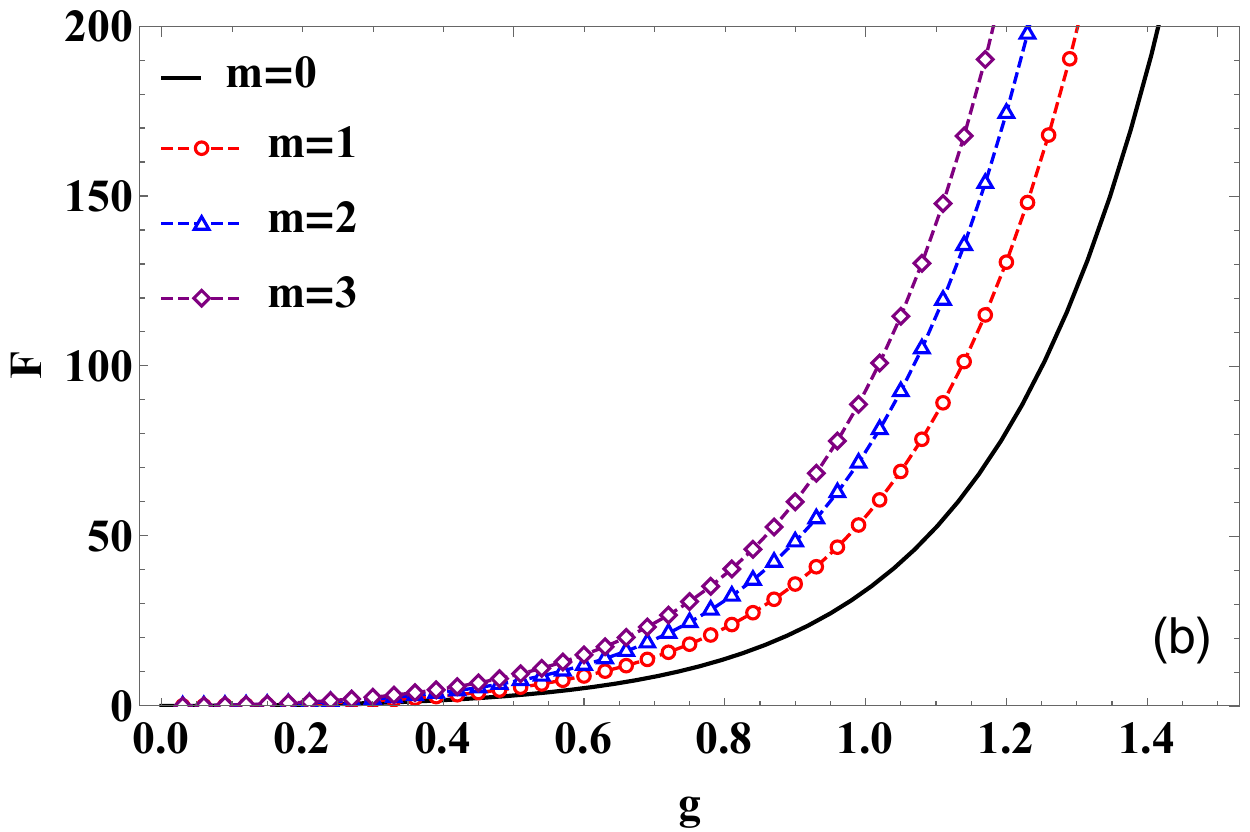}
\caption{The quantum Fisher information F as a function of (a) $\protect%
\beta $ with $g=1$, and $\protect \phi =0.4$, and (b) $g$ with $\protect \beta %
=1$, and $\protect \phi =0.4$.}
\end{figure}

Here, $\left \vert \psi _{\phi }\right \rangle $ represents the state after
undergoing a phase shift. In our model, to account for the influence of the
multiphoton subtraction operations on the ideal QFI, we construct an
equivalent model (see Fig. 6). In this model, we introduce a non-local
operation $(U_{P})$, where $U_{P}=S_{2}^{\dagger }a^{m}S_{2}$. The
multiphoton subtraction operations are incorporated as part of the non-local
operation, and we have $\left \vert \psi _{\phi }\right \rangle
=N_{1}U_{P}U_{\phi }S_{1}\left \vert \psi \right \rangle _{in},$ with $%
\left
\vert \psi _{\phi }^{^{\prime }}\right \rangle =\partial \left \vert
\psi _{\phi }\right \rangle /\partial \phi $. For simplicity, we provide the
calculated result of the QFI under ideal conditions in Appendix B.

In Fig. 7, we illustrate the variation of the QFI with respect to $\beta $
and $g$ under ideal conditions. It is evident that the QFI increases with
both $\beta $ and $g$. This increase can be attributed to the fact that as $%
\beta $ and $g$ increase, the number of photons within the interferometer
also rises, thereby enhancing the interference effects. The reduction in
system uncertainty facilitates stronger interference effects, enabling the
interferometer to extract more information, which in turn enhances the QFI.
Furthermore, when other parameters are the same, the QFI increases with the
number of the photon subtractions. The magnitude of the QFI is closely
related to the non-classicality of the quantum states. Non-Gaussian
operations such as photon subtraction can enhance the non-classical
properties of the quantum states, including quantum statistics, quantum
entanglement, and quantum coherence, ultimately leading to an increase in
the QFI.

Fig. 8 illustrates the variation of the phase sensitivity $\Delta \phi $ and
$\Delta \phi _{QCRB}$ under ideal conditions as the parameters $\beta $ and $%
g$ increase. It is evident that the phase sensitivity $\Delta \phi $
improves as $\beta $ and $g$ increase, with $\Delta \phi _{QCRB}$ following
a similar trend. The enhancement is more pronounced at lower values of these
parameters. Furthermore, the multiphoton subtraction operations
significantly enhance both the phase sensitivity $\Delta \phi $ and $\Delta
\phi _{QCRB}$. However, since the QCRB represents the theoretical limit of
the parameter estimation accuracy, when $m$ is fixed, the dashed lines of
the same color are consistently below the corresponding solid lines. This
observation indicates that while our method can bring the phase sensitivity $%
\Delta \phi $ close to $\Delta \phi _{QCRB}$ under ideal conditions, it
cannot fully achieve this limit, which is consistent with the principles of
quantum limit theory.
\begin{figure}[tph]
\label{Fig8} \centering \includegraphics[width=0.83\columnwidth]{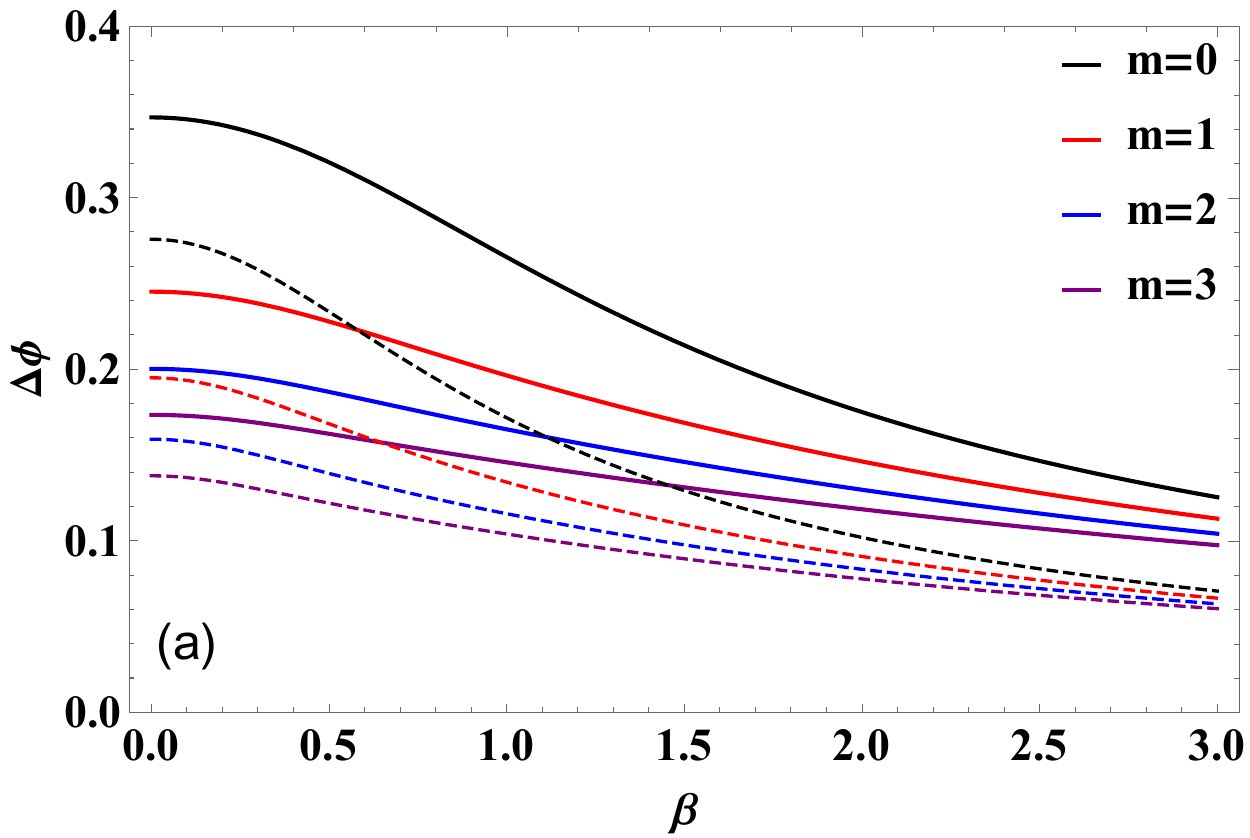} %
\centering \includegraphics[width=0.83\columnwidth]{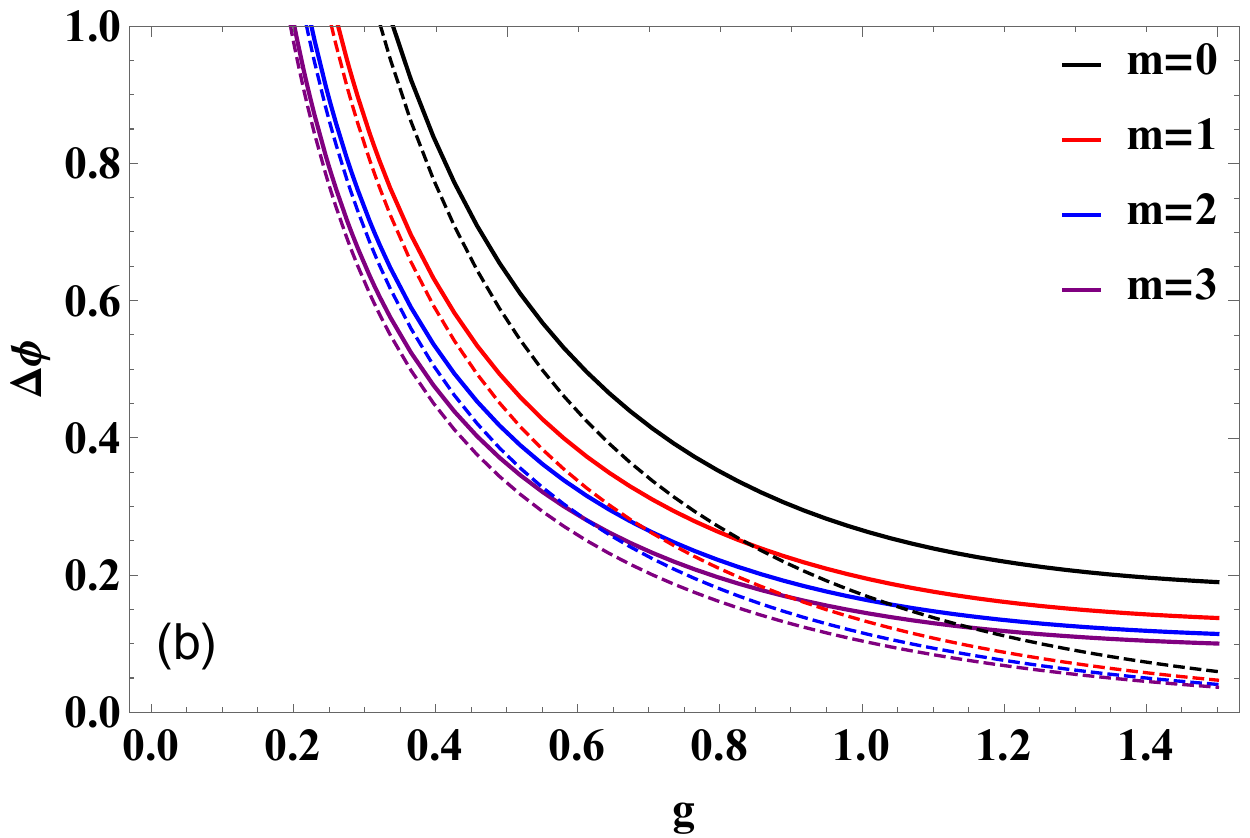}
\caption{The phase sensitivity and QCRB as a function of (a) $\protect \beta $
with $g=1$, and $\protect \phi =0.4$, and (b) $g$ with $\protect \beta =1$,
and $\protect \phi =0.4$. The solid line and dashed line represent the phase
sensitivity $\Delta \protect \phi $ and $\Delta \protect \phi _{QCRB}$,
respectively.}
\end{figure}

\subsection{QFI in the presence of photon losses}

Next, we further explore the influence of photon losses on the QFI. In our
model, the phase shifter induces a minor phase shift within mode $a$. For
simplicity, we concentrate solely on photon losses occurring within mode $a$
inside the interferometer, as shown in Fig. 9. Furthermore, we will compute
the QFI in the presence of photon losses utilizing the method proposed by
Escher $et$ $al.$ \cite{3}. For additional details regarding the
calculation, please refer to Appendix C.
\begin{figure}[tph]
\label{Fig9} \centering \includegraphics[width=1\columnwidth]{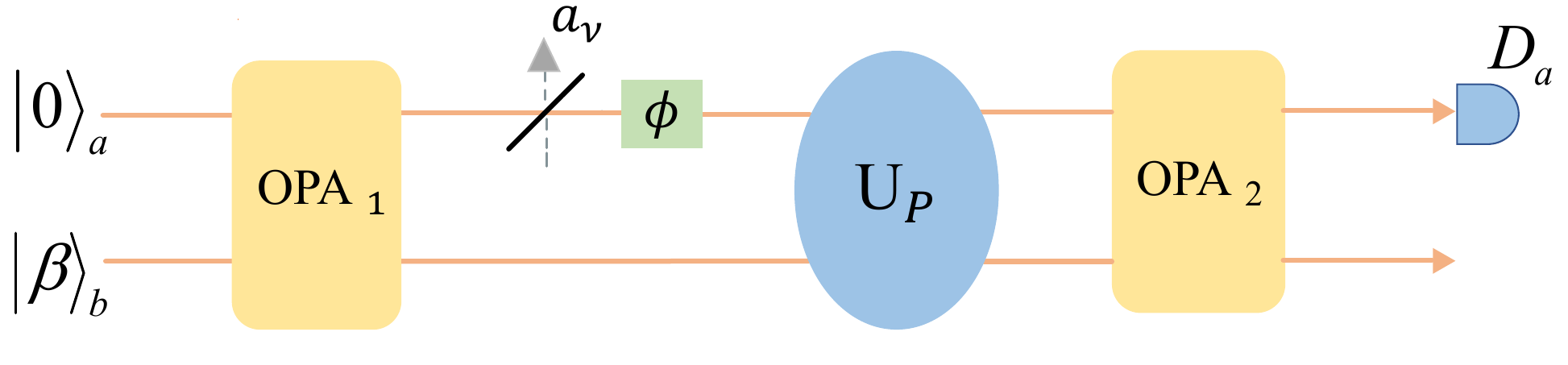}
\caption{{}Schematic diagram of the photon losses on mode $a$, where $U_{P}$
represents a non-local operation, and $U_{P}=S_{2}^{\dagger }a^{m}S_{2}$.}
\end{figure}
\begin{figure}[tph]
\label{Fig10} \centering \includegraphics[width=0.83\columnwidth]{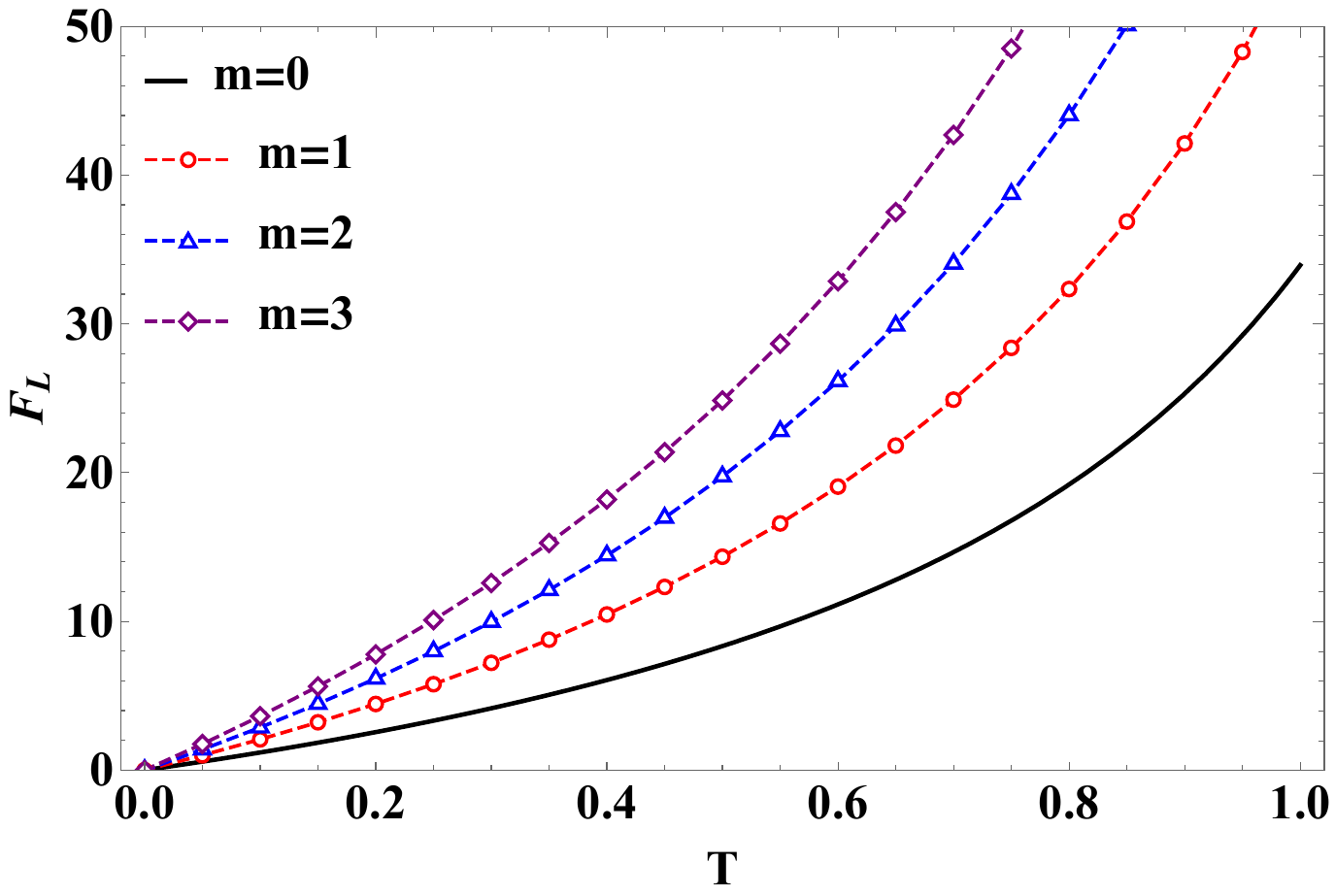}
\caption{The quantum Fisher information F as a function of $T$ with $\protect%
\beta =1$, $g=1$, and $\protect \phi =0.4$.}
\end{figure}
\begin{figure}[tph]
\label{Fig11} \centering \includegraphics[width=0.83\columnwidth]{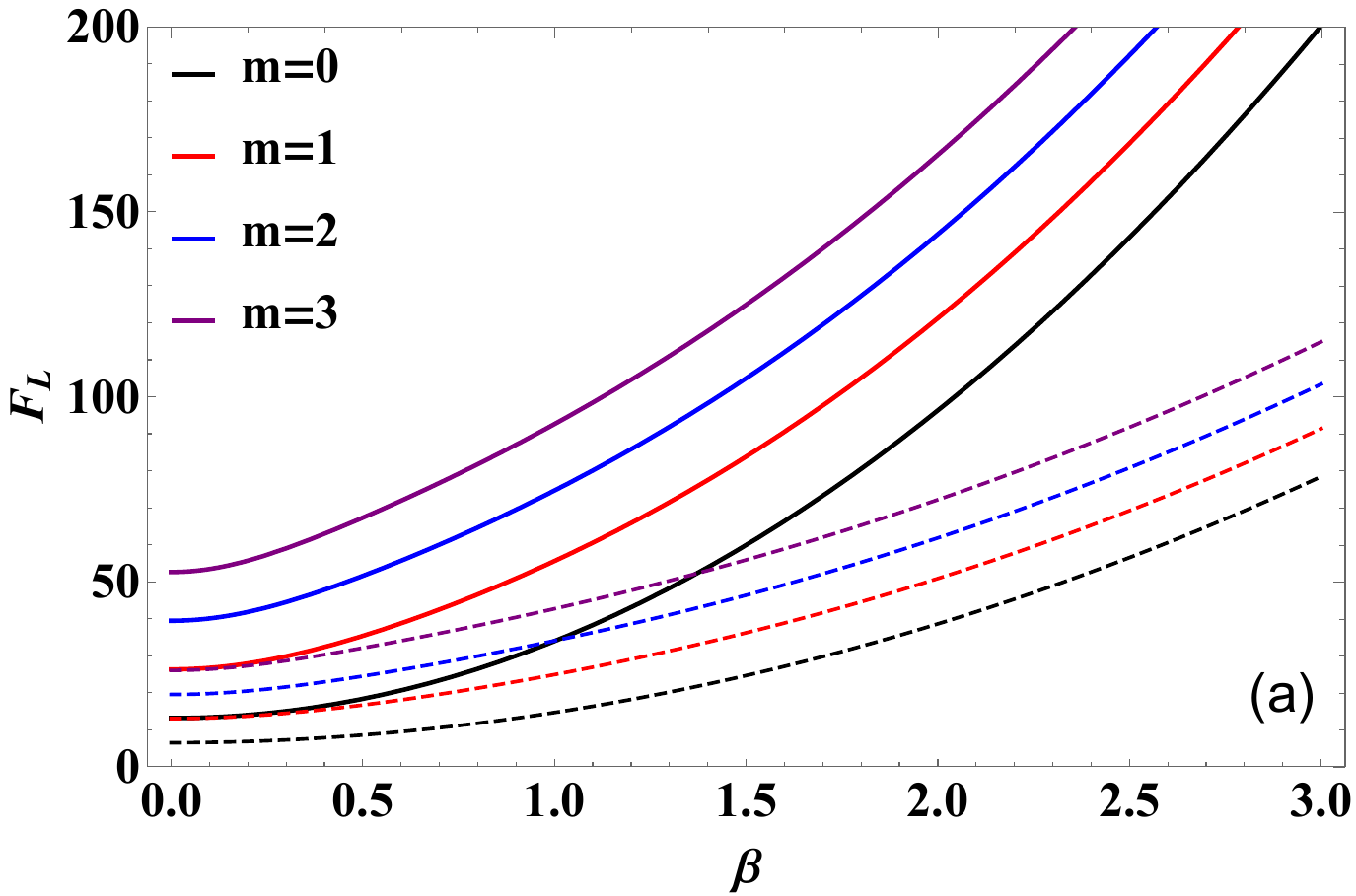}
\centering \includegraphics[width=0.83\columnwidth]{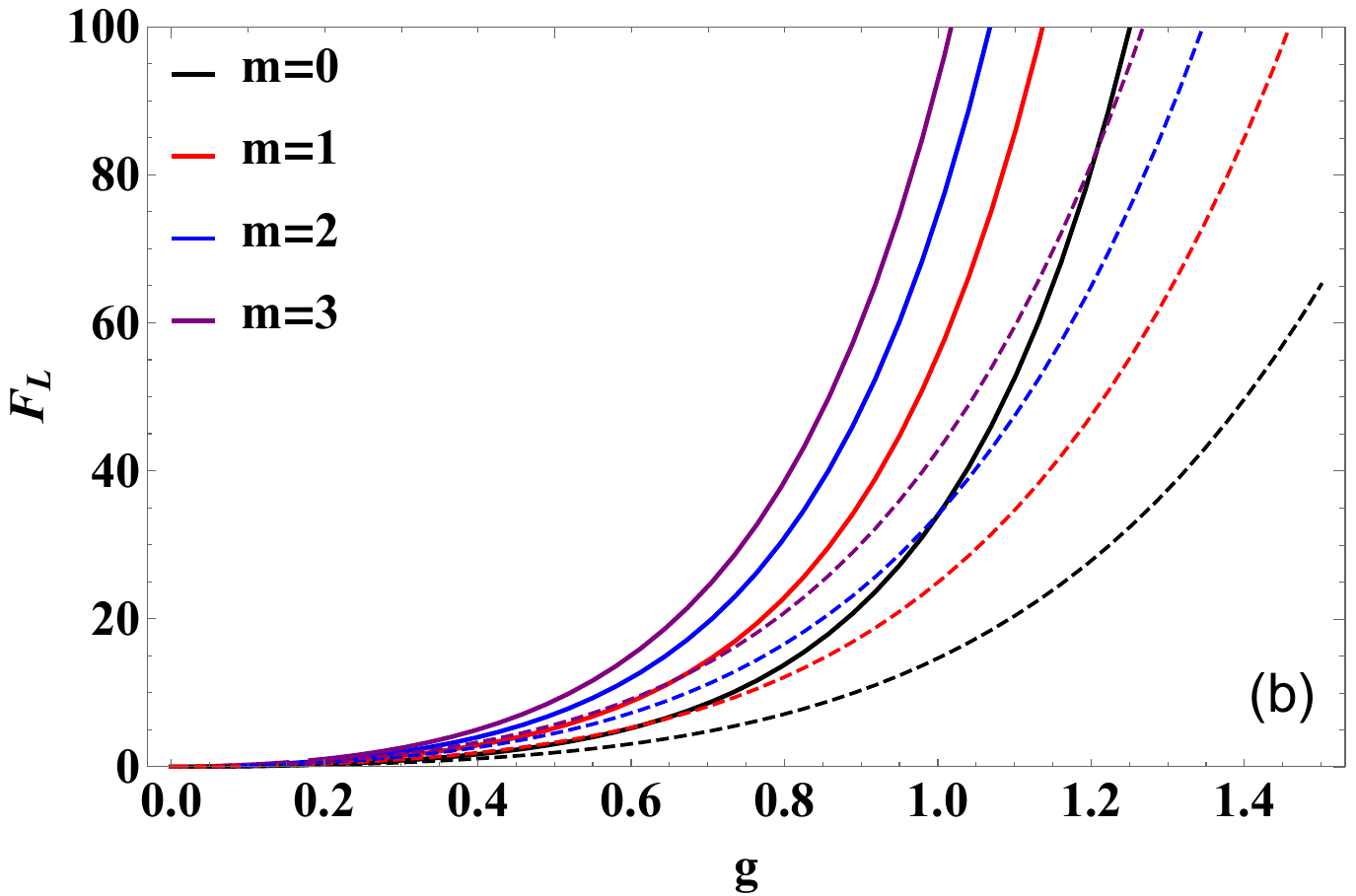}
\caption{The quantum Fisher information F as a function of (a) $\protect%
\beta $ with $g=1$, and $\protect \phi =0.4$, and (b) $g$ with $\protect \beta %
=1$, and $\protect \phi =0.4$. The solid line and dashed line correspond to
the ideal case $\left( T=1\right) $ and the photon losses case $\left(
T=0.7\right) $, respectively.}
\end{figure}

Fig. 10 illustrates the variation of the QFI with the transmissivity $T$ of
the BS that simulates photon losses. It is evident that: (i) photon losses
significantly impact the QFI; as photon losses increase, the QFI
correspondingly decreases. This reduction can be attributed to the
diminished number of photons participating in the interference within the
SU(1,1) interferometer due to these losses, thereby reducing the amount of
information obtainable within the interferometer and ultimately leading to a
decrease in the QFI; (ii) when photon losses remain constant, the QFI
increases with $m$. This indicates that the multiphoton subtraction
operations enhance the non-classicality of the quantum state, improving the
phase sensitivity of the light field and rendering the quantum state more
sensitive to the phase changes. Based on Eq. (\ref{a111}), the QFI
quantitatively describes the sensitivity of a quantum state to phase
changes; thus, the increase in phase sensitivity directly results in a
higher QFI. Therefore, this demonstrates that the multiphoton subtraction
operations can effectively mitigate the impact of internal photon losses in
the interferometer on the QFI.

Meanwhile, Fig. 11 illustrates the variation of the QFI with respect to the
parameters $\beta $ and $g$ under conditions of photon loss. It is evident
that when $\beta $ and $g$ are within a specific range, a $30\%$ photon
losses ($T=0.7$) result in a reduction of over $50\%$ in the QFI,
demonstrating the significant impact of photon losses on the QFI.
Furthermore, in Figs. 11(a) and 11(b), intersections can be observed between
the black solid line and the three dashed lines in red, blue, and purple.
This indicates that the effects of photon losses on the QFI can be mitigated
through the implementation of the multiphoton subtraction operations.

\subsection{Phase sensitivity compared with theoretical limits}

\begin{figure}[tph]
\label{Fig12} \centering \includegraphics[width=0.83\columnwidth]{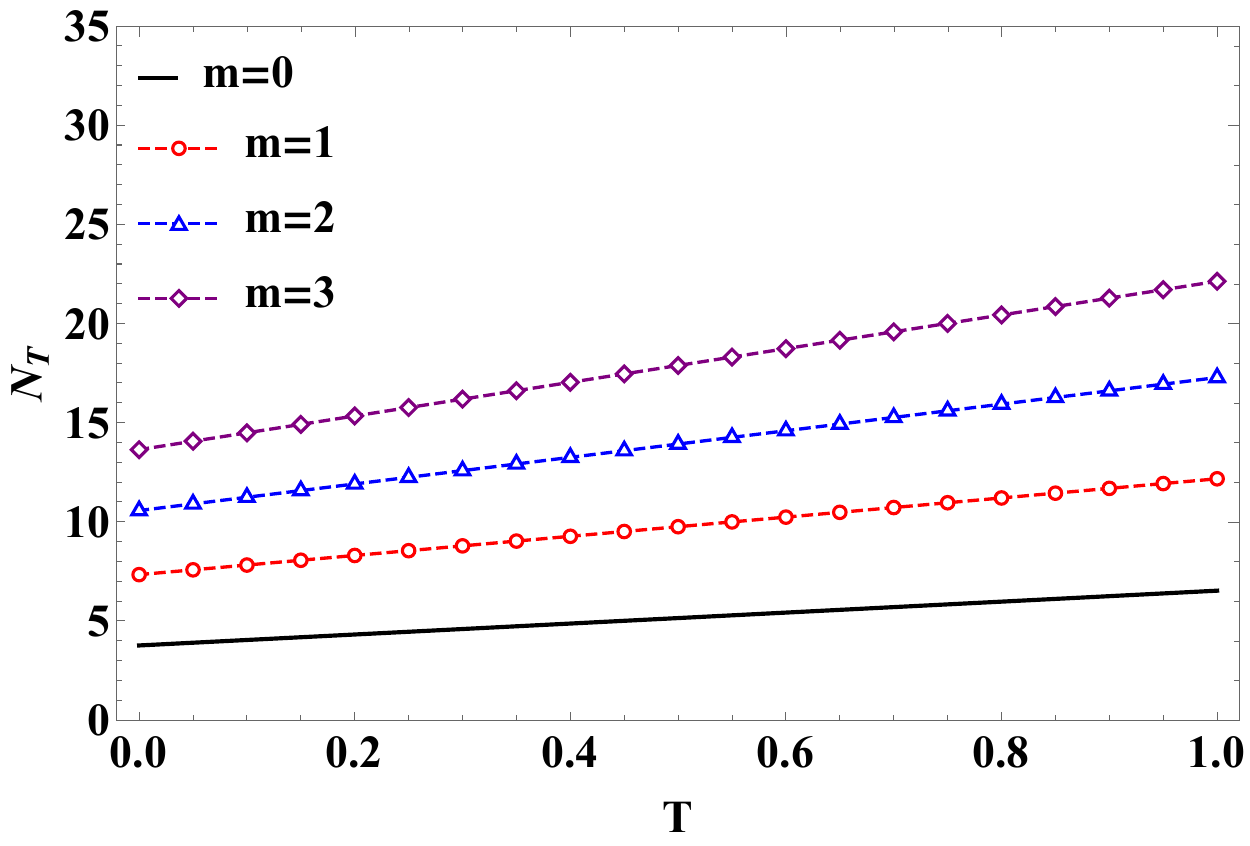}
\caption{The average photon number $N$ as a function of $T$, with $\protect%
\beta =1$, $g=1$, and $\protect \phi =0.4$.}
\end{figure}

In this subsection, we further illustrate the advantages of our approach by
comparing the phase sensitivity of the interferometer with several
theoretical limits, specifically in the presence of internal photon losses.
These limits include the SQL, the Heisenberg limit (HL), and the QCRB.
\begin{figure*}[tph]
\label{Fig13} \centering \includegraphics[width=0.83\columnwidth]{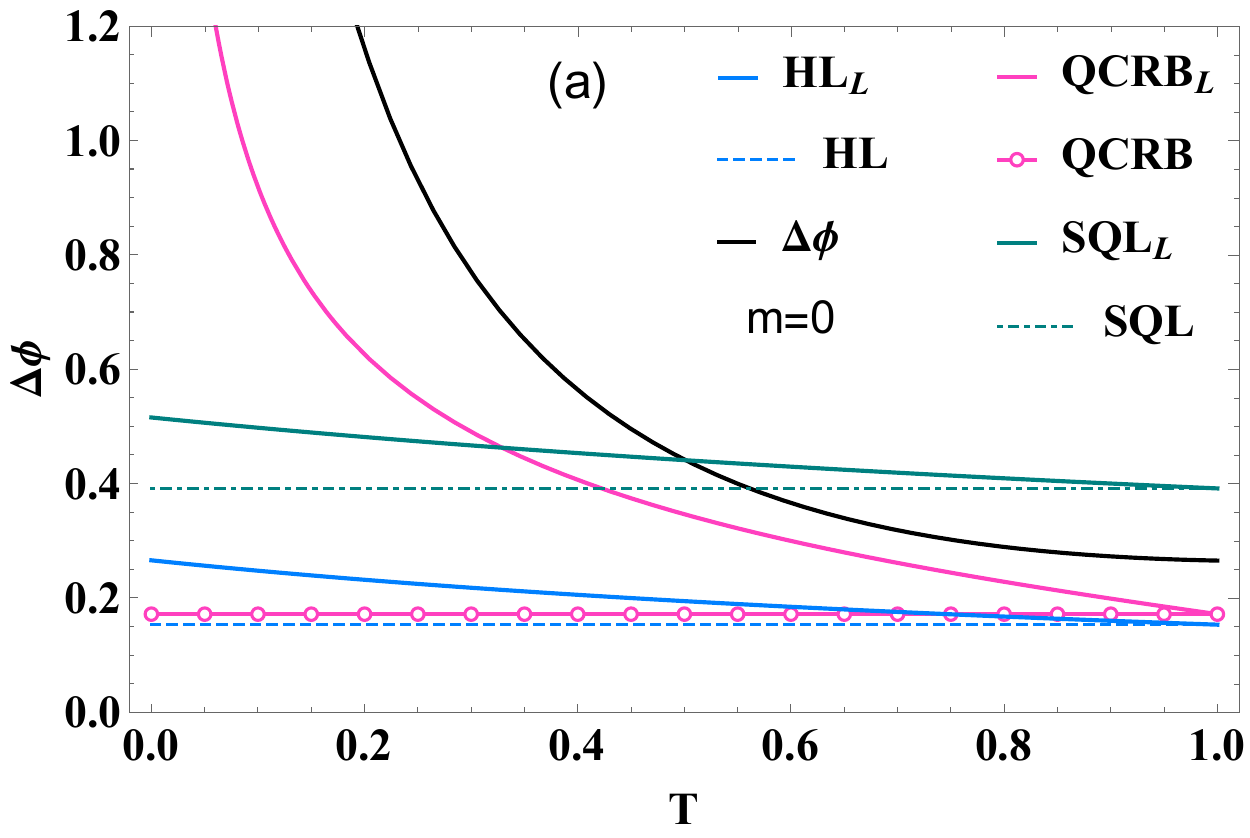}
\centering \includegraphics[width=0.83\columnwidth]{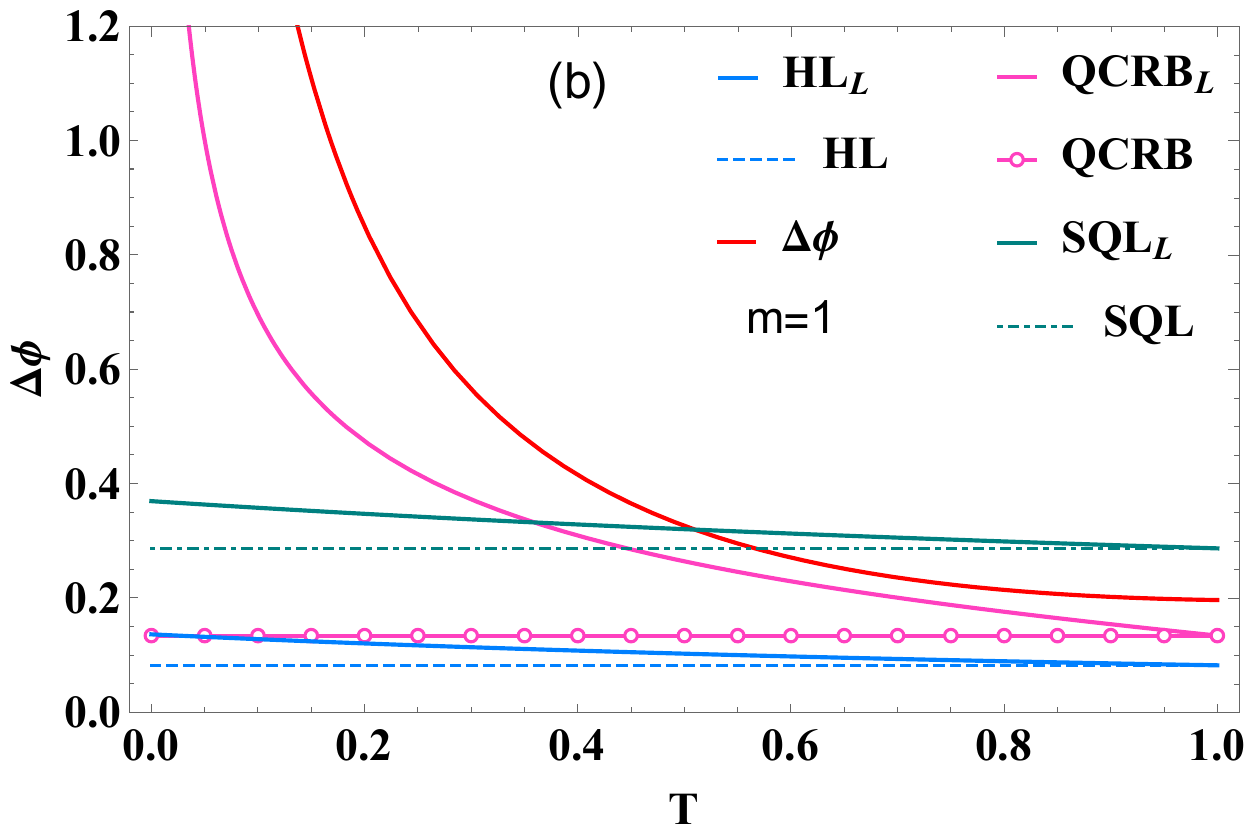} \centering %
\includegraphics[width=0.83\columnwidth]{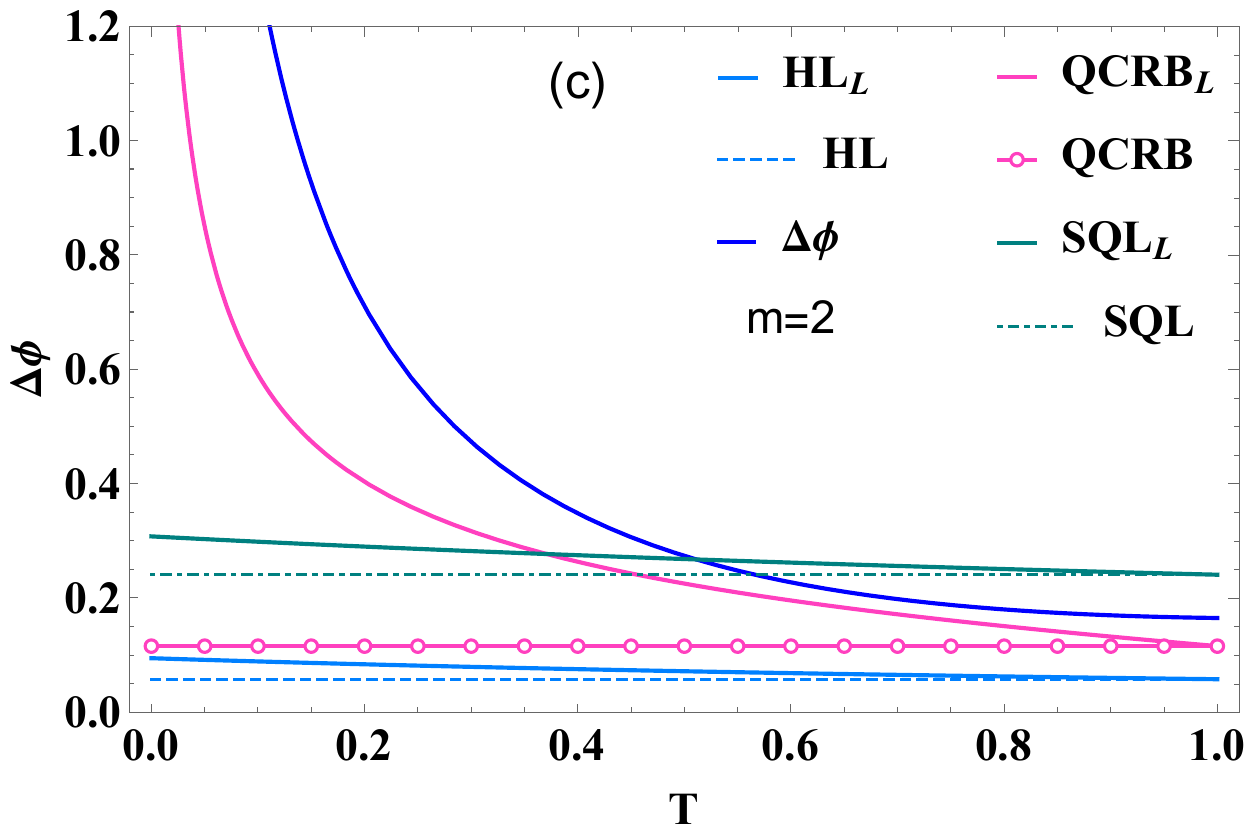} \centering %
\includegraphics[width=0.83\columnwidth]{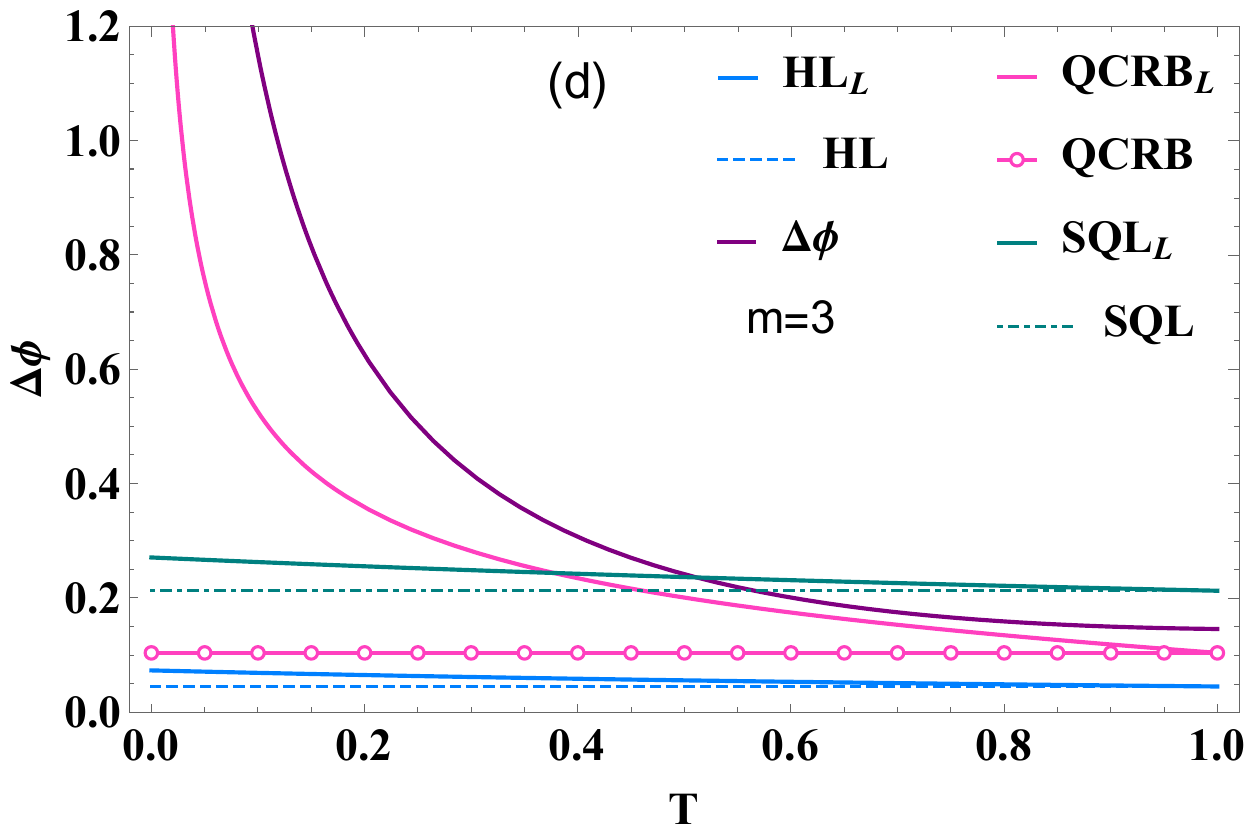}
\caption{The phase sensitivity and theoretical limits as a function of $T$
with $\protect \beta $ $=1$, $g=1$, and $\protect \phi $ $=0.4$. The
subtracting of $0,1,2,$and $3$ photons corresponds to (a), (b), (c), and
(d), respectively.}
\end{figure*}

SQL and HL are associated with the internal mean photon number of the
interferometer $\left( N_{T}\right) $, which can be expressed as%
\begin{eqnarray}
\Delta \phi _{SQL} &=&\frac{1}{\sqrt{N_{T}}},  \label{a9} \\
\Delta \phi _{HL} &=&\frac{1}{N_{T}}.  \label{a10}
\end{eqnarray}

In our model, to account for the effects of the multiphoton subtraction
operations at the output port of the interferometer on the SQL and the HL,
we use the photon count prior to the final OPA in the model shown in Fig. 9
to calculate the average photon number within the interferometer. The
expression for $N_{T}$ is given by

\begin{equation}
N_{T}=\left. _{int}\left \langle \psi \right \vert \left( a^{\dagger
}a+b^{\dagger }b\right) \left \vert \psi \right \rangle _{int}\right. ,
\label{a11}
\end{equation}%
where $\left \vert \psi \right \rangle _{int}=N_{4}U_{P}U_{\phi
}B_{1}S_{1}\left \vert 0\right \rangle _{a}\left \vert \beta \right \rangle
_{b}$, and $N_{4}$ is the normalization coefficient. Further computational
details can be found in Appendix D.

Next, we further examine the impact of the multiphoton subtraction
operations on the average photon number within the interferometer. As shown
in Fig. 12, these operations significantly increase the average photon
number, thereby enhancing the quantity of photons that carry phase
information. This finding contrasts with the conventional understanding that
photon subtraction typically reduces the number of photons. We interpret
photon subtraction as a probabilistic non-Gaussian operation. When the
photon number distribution of the input state is super-Poissonian,
performing photon subtraction leads to an output state characterized by an
increased average photon number. Moreover, in scenarios where photon losses
are relatively low, the multiphoton subtraction operations demonstrate a
more pronounced increase in photon number, suggesting that our approach is
particularly effective under low-loss conditions. Additionally, the SQL and
the HL are correlated with the average number of photons present within the
interferometer. Consequently, the multiphoton subtraction operations enhance
the quantum theoretical limits.

In Fig. 13, we present curves that illustrate the phase sensitivity and the
corresponding theoretical limits for $m=0,1,2$, and $3$ as functions of the
transmittance $T$. It is evident from Fig. 13 that: (i) the multiphoton
subtraction operations enhance the phase sensitivity while simultaneously
improving the theoretical precision limit. Our scheme demonstrates that the
phase sensitivity can surpass the SQL even when photon losses reach $40\%$,
highlighting the robustness of the approach; (ii) photon losses in mode $a$
within the interferometer significantly impact the QCRB but have a lesser
effect on the SQL and the HL; (iii) in our scheme, the HL precision exceeds
the QCRB precision under ideal conditions. Furthermore, as $m$ increases,
the phase sensitivity approaches the QCRB more closely in the presence of
photon losses.

\section{Conclusion}

In this paper, we investigate the effects of the multiphoton subtraction
operations at the output port of the SU(1,1) interferometer on the phase
sensitivity. Our study utilizes vacuum and coherent states as the inputs,
employing intensity detection as the measurement method. We first explore
the impact of various orders of the photon subtraction operations on the
phase sensitivity under ideal conditions, revealing that an increase in the
number of photon subtractions gradually enhances the phase sensitivity.
Additionally, we compare the phase sensitivity obtained from detections at
the $a$ and $b$ ports of the interferometer, finding that detection at the $%
a $ port is more effective.

We also examine the effects of photon losses on the phase sensitivity and
the QFI. Our analysis reveals that internal photon losses in the SU(1,1)
interferometer have a more pronounced impact on the phase sensitivity
compared to external losses. Furthermore, to assess the impact of the photon
subtraction operations at the output port on the QFI, we propose an
equivalent model that treats the multiphoton subtraction operations as part
of a non-local operation. The results indicate that even under relatively
severe photon losses, our method significantly improves both the phase
sensitivity and the QFI. Finally, we compare the phase sensitivity with the
theoretical limits under conditions of photon loss, demonstrating that our
approach enables the phase sensitivity to surpass the SQL while also
approaching the HL and the QCRB even with $40\%$ photon losses. Overall, our
research provides theoretical support for the advancement of quantum
information science.

\begin{acknowledgments}
This work is supported by the National Natural Science Foundation of China (Grants No. 11964013 and No. 12104195) and the Jiangxi Provincial Natural Science Foundation (Grants No. 20242BAB26009 and 20232BAB211033), Jiangxi Provincial Key Laboratory of Advanced Electronic Materials and Devices (Grant No. 2024SSY03011), as well as Jiangxi Civil-Military Integration Research Institute (Grant No. 2024JXRH0Y07).
\end{acknowledgments}\bigskip

\bigskip \textbf{APPENDIX\ A : PHASE SENSITIVITY IN IDEAL CASE}

In this Appendix, we provide the formula for the phase sensitivity under
ideal conditions. By following a series of operations, the output state can
be expressed as follows

\begin{equation}
\left \vert \psi \right \rangle _{out}=N_{1}a^{m}S_{2}U_{\phi }S_{1}\left
\vert \psi \right \rangle _{in}.  \tag{A1}
\end{equation}

Before deriving the phase sensitivity, we first introduce a formula, i.e.,

\begin{align}
& S_{1}^{\dag }U_{\phi }^{\dagger }S_{2}^{\dagger }a^{\dagger
m}a^{m}S_{2}U_{\phi }S_{1}  \notag \\
& =\frac{\partial ^{2m}}{\partial t^{m}\partial s^{m}}[S_{1}^{\dag }U_{\phi
}^{\dagger }S_{2}^{\dagger }\exp \left( ta^{\dagger }\right)  \notag \\
& \times \exp \left( sa\right) S_{2}U_{\phi }S_{1}]|_{t=s=0}  \notag \\
& =\frac{\partial ^{2m}}{\partial t^{m}\partial s^{m}}[\exp \left(
tw_{2}a^{\dagger }\right) \exp \left( tw_{1}b\right)  \notag \\
& \times \exp \left( sw_{2}^{\ast }a\right) \exp \left( sw_{1}^{\ast
}b^{\dagger }\right) ]|_{t=s=0},  \tag{A2}
\end{align}%
where we have set%
\begin{align}
w_{1}& =\frac{1}{2}\sinh 2g\left( 1-e^{-i\phi }\right) ,  \tag{A3} \\
w_{2}& =\cosh ^{2}ge^{-i\phi }-\sinh ^{2}g,  \tag{A4}
\end{align}

According to the normalization condition, we have%
\begin{align}
& \left. _{out}\left \langle \psi |\psi \right \rangle _{out}\right.  \notag
\\
& =N_{1}^{2}\left. _{in}\left \langle \psi \right \vert S_{1}^{\dagger
}U_{\phi }^{\dagger }S_{2}^{\dagger }a^{\dagger m}a^{m}S_{2}U_{\phi
}S_{1}\left \vert \psi \right \rangle _{in}\right.  \notag \\
& =N_{1}^{2}\frac{\partial ^{2m}}{\partial t^{m}\partial s^{m}}\{ \exp \left[
t\left( w_{2}a^{\dagger }+w_{1}b\right) \right]  \notag \\
& \times \exp \left[ s\left( w_{2}^{\ast }a+w_{1}^{\ast }b^{\dagger }\right) %
\right] \}|_{t=s=0}  \notag \\
& =N_{1}^{2}G_{m}e^{A_{1}}  \notag \\
& =1,  \tag{A5}
\end{align}%
where%
\begin{equation}
G_{m}e^{\left( \cdot \right) }=\frac{\partial ^{2m}}{\partial t^{m}\partial
s^{m}}e^{\left( \cdot \right) }|_{t=s=0},  \tag{A6}
\end{equation}%
and%
\begin{equation}
A_{1}=st\left \vert w_{1}\right \vert ^{2}+\left( tw_{1}+sw_{1}^{\ast
}\right) \beta ,  \tag{A7}
\end{equation}%
\begin{equation}
N_{1}=\left( G_{m}e^{A_{1}}\right) ^{-\frac{1}{2}}.  \tag{A8}
\end{equation}%
Here, $m$ is an integer representing the number of the photon subtractions,
while $s$ and $t$ are differential variables. After differentiation, these
variables all become $0$.

Then, based on Eqs. (A1)-(A8), the phase sensitivity in the ideal case is
given by

\begin{equation}
\Delta ^{2}\phi =\frac{\left \langle N^{2}\right \rangle -\left \langle
N\right \rangle ^{2}}{\left \vert \partial _{\phi }\left \langle N\right
\rangle \right \vert ^{2}},  \tag{A9}
\end{equation}%
where%
\begin{equation}
\left \langle N\right \rangle =_{out}\left \langle \psi \right \vert
a^{\dagger }a\left \vert \psi \right \rangle
_{out}=N_{1}^{2}G_{m+1}e^{A_{1}},  \tag{A10}
\end{equation}%
and%
\begin{align}
\left \langle N^{2}\right \rangle & =_{out}\left \langle \psi \right \vert
(a^{\dagger }a)^{2}\left \vert \psi \right \rangle _{out}  \notag \\
& =N_{1}^{2}\left( G_{m+1}e^{A_{1}}+G_{m+2}e^{A_{1}}\right) .  \tag{A11}
\end{align}

\textbf{APPENDIX\ B : QFI UNDER IDEAL CONDITION}

\bigskip To calculate the ideal QFI, we introduce an equivalent model shown
in Fig. 6. According to Eq. (\ref{a111}), we have $\left \vert \psi _{\phi
}\right \rangle =N_{1}U_{P}U_{\phi }S_{1}\left \vert 0\right \rangle
_{a}\otimes \left \vert \beta \right \rangle _{b}$, with $%
U_{P}=S_{2}^{\dagger }a^{m}S_{2}$, and derived $\left \vert \psi _{\phi
}^{^{\prime }}\right \rangle =\frac{\partial \left \vert \psi _{\phi
}\right
\rangle }{\partial \phi }$. Then, the QFI in the ideal case is
given by%
\begin{equation}
F=4\left[ \left \langle \psi _{\phi }^{^{\prime }}|\psi _{\phi }^{^{\prime
}}\right \rangle -\left \vert \left \langle \psi _{\phi }^{^{\prime }}|\psi
_{\phi }\right \rangle \right \vert ^{2}\right] ,  \tag{B1}
\end{equation}%
with%
\begin{align}
\left \langle \psi _{\phi }^{^{\prime }}|\psi _{\phi }^{^{\prime }}\right
\rangle & =N_{1}^{2}Y_{m}e^{F_{1}}+\left( \frac{\partial N_{1}}{\partial
\phi }\right) ^{2}G_{m}e^{F_{2}}  \notag \\
& +iN_{1}\frac{\partial N_{1}}{\partial \phi }H_{m}e^{F_{4}}-iN_{1}\frac{%
\partial N_{1}}{\partial \phi }H_{m}e^{F_{3}},  \tag{B2} \\
\left \langle \psi _{\phi }^{^{\prime }}|\psi _{\phi }\right \rangle &
=-iN_{1}^{2}H_{m}e^{F_{3}}+\frac{\partial N_{1}}{\partial \phi }%
N_{1}G_{m}e^{F_{2}},  \tag{B3} \\
\left \langle \psi _{\phi }|\psi _{\phi }^{\prime }\right \rangle &
=iN_{1}^{2}H_{m}e^{F_{4}}+\frac{\partial N_{1}}{\partial \phi }%
N_{1}G_{m}e^{F_{2}},  \tag{B4}
\end{align}%
where we have defined

\begin{align}
H_{m}e^{\left( \cdot \right) }& =\frac{\partial ^{2m+2}}{\partial
t^{m}\partial s^{m}\partial c\partial d}e^{\left( \cdot \right)
}|_{t=s=c=d=0},  \tag{B5} \\
Y_{m}e^{\left( \cdot \right) }& =\frac{\partial ^{2m+4}}{\partial
t^{m}\partial s^{m}\partial c\partial d\partial p\partial h}e^{\left( \cdot
\right) }|_{t=s=c=d=p=h=0}.  \tag{B6}
\end{align}%
Here, $m$ is an integer representing the number of the photon subtractions,
while $s$, $t$, $c$, $d$, $p$, and $h$ are differential variables. After
differentiation, these variables all become $0$.

In addition, we have set%
\begin{align}
F_{1}& =df_{1}^{\ast }\left( tf_{3}+pf_{1}\right) +spf_{1}f_{3}^{\ast
}+ph\left \vert f_{2}\right \vert ^{2}  \notag \\
& +tf_{4}\left( sf_{4}^{\ast }+hf_{2}^{\ast }\right) +cf_{2}\left(
sf_{4}^{\ast }+hf_{2}^{\ast }+df_{2}^{\ast }\right)  \notag \\
& +\left( sf_{4}^{\ast }+hf_{2}^{\ast }+df_{2}^{\ast
}+cf_{2}+tf_{4}+pf_{2}\right) \beta ,  \tag{B7} \\
F_{2}& =st\left \vert f_{4}\right \vert ^{2}+\left( sf_{4}^{\ast
}+tf_{4}\right) \beta ,  \tag{B8} \\
F_{3}& =tdf_{1}^{\ast }f_{3}+cd\left \vert f_{2}\right \vert ^{2}+st\left
\vert f_{4}\right \vert ^{2}+csf_{2}f_{4}^{\ast }  \notag \\
& +\left( df_{2}^{\ast }+sf_{4}^{\ast }+cf_{2}+tf_{4}\right) \beta ,
\tag{B9} \\
F_{4}& =csf_{1}f_{3}^{\ast }+st\left \vert f_{4}\right \vert ^{2}+cd\left
\vert f_{2}\right \vert ^{2}+tdf_{2}^{\ast }f_{4}  \notag \\
& +\left( sf_{4}^{\ast }+df_{2}^{\ast }+tf_{4}+cf_{2}\right) \beta ,
\tag{B10}
\end{align}%
with%
\begin{align}
f_{1}& =\cosh ge^{-i\phi },  \notag \\
f_{2}& =-\sinh ge^{-i\phi },  \notag \\
f_{3}& =\cosh ^{2}ge^{-i\phi }-\sinh ^{2}g,  \notag \\
f_{4}& =\frac{1}{2}\sinh 2g(1-e^{-i\phi }).  \tag{B11}
\end{align}

\textbf{APPENDIX\ C :} \bigskip \textbf{QFI in the presence of photon losses}
\begin{figure}[tph]
\label{Fig14} \centering \includegraphics[width=1\columnwidth]{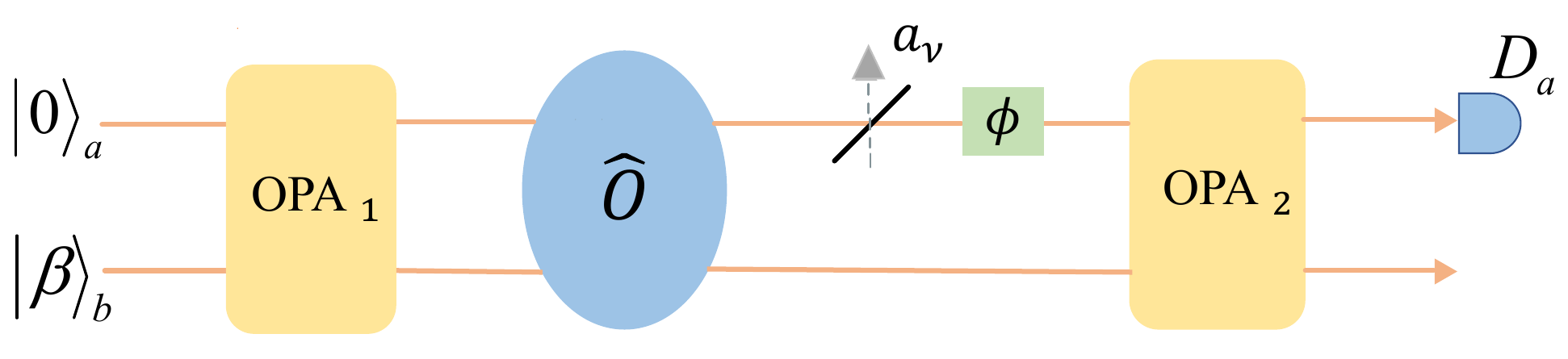}
\caption{{}Schematic diagram of the photon losses on mode $a$, where $\hat{O}
$ represents a non-local operation, and $\hat{O}=\frac{\partial ^{m}}{%
\partial s^{m}}\exp \left[ s\left( ae^{i\protect \phi }\protect \sqrt{\protect%
\eta }\cosh g+b^{\dagger }\sinh g\right) \right] |_{s=0}$.}
\end{figure}

In this section, we further derive the QFI in the presence of photon losses
within the system, as shown in Fig. 9. It is crucial to emphasize that for
the SU(1,1) interferometer, the QFI quantifies the amount of information
acquired before the final OPA. Consequently, we focus on the quantum state
before the second OPA. We denote the initial probe state of the SU(1,1)
interferometer system $S$ as $\left \vert \psi \right \rangle
_{S}=S_{1}\left \vert 0\right \rangle _{a}\otimes \left \vert \beta
\right
\rangle _{b}.$ Due to photon losses in the system, the process of
encoding the unknown phase into the probe state $\left \vert \psi
\right
\rangle _{S}$ is no longer a unitary evolution. Therefore, we
transform the problem into parameter estimation under unitary evolution in
an extended system $S+E$. In this extended system, the probe state $%
\left
\vert \psi \right \rangle _{S}$ under unitary evolution $%
U_{S+E}\left( \phi \right) $ is given by \cite{3}

\begin{align}
\left \vert \psi \right \rangle _{S+E}& =N_{3}U_{P}U_{S+E}\left( \phi
\right) \left \vert \psi \right \rangle _{S}\left \vert 0\right \rangle _{E}
\notag \\
& =N_{3}U_{S+E}\left( \phi \right) \hat{O}\left \vert \psi \right \rangle
_{S}\left \vert 0\right \rangle _{E}  \notag \\
& =N_{3}\overset{\infty }{\underset{l=0}{\sum }}\Pi _{l}\left( \phi \right)
\hat{O}\left \vert \psi \right \rangle _{S}\left \vert l\right \rangle _{E}
\notag \\
& =\overset{\infty }{\underset{l=0}{\sum }}\Pi _{l}\left( \phi \right) \left
\vert \Psi \right \rangle \left \vert l\right \rangle _{E},  \tag{C1}
\end{align}%
where $\left \vert \Psi \right \rangle =N_{3}\hat{O}\left \vert \psi
\right
\rangle _{S},$ $N_{3}$ is the normalization factor, while $\hat{O}$
represents the equivalent operator that transforms the non-local operation ($%
U_{P}$) into a form preceding the photon losses, with its expression given
by $\hat{O}=\frac{\partial ^{m}}{\partial s^{m}}\exp \left[ s\left(
ae^{i\phi }\sqrt{\eta }\cosh g+b^{\dagger }\sinh g\right) \right] |_{s=0}$
(as shown in Fig. 14). Here, $\left \vert 0\right \rangle _{E}$ represents
the initial state in an environment with photon losses, and $\left \vert
l\right \rangle _{E}$ denotes the orthogonal basis corresponding to $%
\left
\vert 0\right \rangle _{E}.$ Additionally, $\Pi _{l}\left( \phi
\right) $ is the Kraus operator used to describe photon losses, and its form
is given by \cite{3}

\begin{equation}
\Pi _{l}\left( \phi \right) =\sqrt{\frac{\left( 1-\eta \right) ^{l}}{l!}}%
e^{i\phi \left( n-\alpha l\right) }\eta ^{\frac{n}{2}}a^{l},  \tag{C2}
\end{equation}%
where, $e^{i\phi n}$ represents the phase shifter operator. We introduce the
parameter $\alpha $, where $\alpha =0$ corresponding to photon losses before
the phase shifter, and $\alpha =-1$ corresponding to photon losses after the
phase shifter. The parameter $\eta $ represents the transmissivity of the BS
that simulating photon losses, with $\eta =1$ corresponding to a lossless
condition.

In the extended system $S+E$, the QFI in the presece of photon losses can be
expressed as \cite{3}

\begin{equation}
F_{L}=C_{Q}\left[ \left \vert \psi \right \rangle _{S},\Pi _{l}\left( \phi
\right) \right] _{\min }.  \tag{C3}
\end{equation}

In this extended system, the upper bound of the QFI is given by \cite{3}%
\begin{align}
& C_{Q}\left[ \left \vert \psi \right \rangle _{S},\Pi _{l}\left( \phi
\right) \right]  \notag \\
& =4\left[ _{S+E}\left \langle \psi ^{\prime }|\psi ^{\prime }\right \rangle
_{S+E}-\left \vert _{S+E}\left \langle \psi ^{\prime }|\psi \right \rangle
_{S+E}\right \vert ^{2}\right] .  \tag{C4}
\end{align}

Based on Eqs. (C1) and (C4), $C_{Q}\left[ \left \vert \psi \right \rangle
_{S},\Pi _{l}\left( \phi \right) \right] $ can be expressed as

\begin{equation}
C_{Q}\left[ \left \vert \psi \right \rangle _{S},\Pi _{l}\left( \phi \right) %
\right] =4\left[ \left \langle h_{1}\right \rangle -\left \vert \left
\langle h_{2}\right \rangle \right \vert ^{2}\right] ,  \tag{C5}
\end{equation}%
where $\left \langle \cdot \right \rangle =\left. _{S}\left \langle \psi
\right \vert \cdot \left \vert \psi \right \rangle _{S}\right. $, and

\begin{align}
h_{1}& =\overset{\infty }{\underset{l=0}{\sum }}\frac{dN_{3}\hat{O}^{\dagger
}\Pi _{l}^{\dagger }\left( \phi \right) }{d\phi }\frac{dN_{3}\Pi _{l}\left(
\phi \right) \hat{O}}{d\phi }  \notag \\
& =\frac{dN_{3}\hat{O}^{\dagger }}{d\phi }\frac{dN_{3}\hat{O}}{d\phi }+N_{3}%
\hat{O}^{\dagger }H_{1}N_{3}\hat{O}  \notag \\
& +\frac{dN_{3}\hat{O}^{\dagger }}{d\phi }iH_{2}^{\dagger }N_{3}\hat{O}+N_{3}%
\hat{O}^{\dagger }\left( -iH_{2}\right) \frac{dN_{3}\hat{O}}{d\phi },
\tag{C6} \\
h_{2}& =i\left. \overset{\infty }{\underset{l=0}{\sum }}\frac{dN_{3}\hat{O}%
^{\dagger }\Pi _{l}^{\dagger }\left( \phi \right) }{d\phi }\Pi _{l}\left(
\phi \right) N_{3}\hat{O}\right.  \notag \\
& =i\frac{dN_{3}\hat{O}^{\dagger }}{d\phi }N_{3}\hat{O}+N_{3}\hat{O}%
^{\dagger }H_{2}N_{3}\hat{O}.  \tag{C7}
\end{align}%
where%
\begin{align}
H_{1}& =\overset{\infty }{\underset{l=0}{\sum }}\frac{d\Pi _{l}^{\dagger
}\left( \phi \right) }{d\phi }\frac{d\Pi _{l}\left( \phi \right) }{d\phi },
\tag{C8} \\
H_{2}& =i\overset{\infty }{\underset{l=0}{\sum }}\frac{d\Pi _{l}^{\dagger
}\left( \phi \right) }{d\phi }\Pi _{l}\left( \phi \right) .  \tag{C9}
\end{align}%
with $H_{1}$, $H_{2}$ being expressed by \cite{3}
\begin{align}
H_{1}& =\left[ 1-\left( 1+\alpha \right) \left( 1-\eta \right) \right]
^{2}n^{2}  \notag \\
& +\left( 1+\alpha \right) ^{2}\eta \left( 1-\eta \right) n,  \tag{C10} \\
H_{2}& =\left[ 1-\left( 1+\alpha \right) \left( 1-\eta \right) \right] n.
\tag{C11}
\end{align}

Based on Eqs. (C2) and (C5)-(C9), $C_{Q}\left[ \left \vert \psi
\right
\rangle _{S},\Pi _{l}\left( \phi \right) \right] $ is given by

\begin{align}
& C_{Q}\left[ \left \vert \psi \right \rangle _{S},\Pi _{l}\left( \phi
\right) \right]  \notag \\
& =4\left[ \left \langle h_{1}\right \rangle -\left \vert \left \langle
h_{2}\right \rangle \right \vert ^{2}\right]  \notag \\
& =4\left. _{S}\left \langle \psi \right \vert \frac{dN_{3}\hat{O}^{\dagger }%
}{d\phi }\frac{dN_{3}\hat{O}}{d\phi }\left \vert \psi \right \rangle
_{S}\right.  \notag \\
& +4\left. _{S}\left \langle \psi \right \vert N_{3}\hat{O}^{\dagger
}H_{1}N_{3}\hat{O}\left \vert \psi \right \rangle _{S}\right.  \notag \\
& +4\left. _{S}\left \langle \psi \right \vert \frac{dN_{3}\hat{O}^{\dagger }%
}{d\phi }iH_{2}^{\dagger }N_{3}\hat{O}\left \vert \psi \right \rangle
_{S}\right.  \notag \\
& +4\left. _{S}\left \langle \psi \right \vert N_{3}\hat{O}^{\dagger }\left(
-iH_{2}\right) \frac{dN_{3}\hat{O}}{d\phi }\left \vert \psi \right \rangle
_{S}\right.  \notag \\
& -4\left \vert _{S}\left \langle \psi \right \vert i\frac{dN_{3}\hat{O}%
^{\dagger }}{d\phi }N_{3}\hat{O}+N_{3}^{2}\hat{O}^{\dagger }H_{2}\hat{O}%
\left \vert \psi \right \rangle _{S}\right \vert ^{2}  \notag \\
& =4\left. \left \langle \tilde{\Psi}|\tilde{\Psi}\right \rangle \right.
+4\left \langle \Psi \right \vert H_{1}\left \vert \Psi \right \rangle
\notag \\
& +4\left \langle \tilde{\Psi}\right \vert iH_{2}^{\dagger }\left \vert \Psi
\right \rangle -4\left \langle \Psi \right \vert iH_{2}\left \vert \tilde{%
\Psi}\right \rangle  \notag \\
& -4\left \vert \left( i\left \langle \tilde{\Psi}|\Psi \right \rangle
+\left \langle \Psi \right \vert H_{2}\left \vert \Psi \right \rangle
\right) \right \vert ^{2},  \tag{C12}
\end{align}%
where $\left \vert \tilde{\Psi}\right \rangle =\partial \left \vert \Psi
\right \rangle /\partial \phi $, and $n=a^{\dagger }a.$

To obtain the $C_{Q}\left[ \left \vert \psi \right \rangle _{S},\Pi
_{l}\left( \phi \right) \right] _{\min }$, we set $\partial C_{Q}\left[
\left \vert \psi \right \rangle _{S},\Pi _{l}\left( \phi \right) \right]
/\partial \alpha =0.$ The expression for the QFI under photon losses is
given by%
\begin{align}
F_{L}& =C_{Q}\left[ \left \vert \psi \right \rangle _{S},\Pi _{l}\left( \phi
\right) \right] _{\min }  \notag \\
& =4\left[ \left \langle \tilde{\Psi}|\tilde{\Psi}\right \rangle -\left
\vert \left \langle \tilde{\Psi}|\Psi \right \rangle \right \vert ^{2}\right]
\notag \\
& +\frac{1}{\left[ \left( 1-\eta \right) \left \langle \Psi \right \vert
\Delta n_{a}^{2}\left \vert \Psi \right \rangle +\eta \left \langle \Psi
\right \vert n\left \vert \Psi \right \rangle \right] }  \notag \\
& \times \lbrack 4\eta \left( \left \langle \Psi \right \vert n\left \vert
\Psi \right \rangle \right) \times (\left \langle \Psi \right \vert \Delta
n_{a}^{2}\left \vert \Psi \right \rangle  \notag \\
& +i\left \langle \Psi \right \vert n\left \vert \Psi \right \rangle \left
\langle \Psi |\tilde{\Psi}\right \rangle -i\left \langle \Psi \right \vert
n\left \vert \Psi \right \rangle \left \langle \tilde{\Psi}|\Psi \right
\rangle )  \notag \\
& +i\left \langle \tilde{\Psi}\right \vert n\left \vert \Psi \right \rangle
-i\left \langle \Psi \right \vert n\left \vert \tilde{\Psi}\right \rangle
\notag \\
& +\left( 1-\eta \right) (\left \langle \tilde{\Psi}\right \vert n\left
\vert \Psi \right \rangle -\left \langle \Psi \right \vert n\left \vert
\tilde{\Psi}\right \rangle  \notag \\
& +\left \langle \Psi \right \vert n\left \vert \Psi \right \rangle \left
\langle \Psi |\tilde{\Psi}\right \rangle -\left \langle \Psi \right \vert
n\left \vert \Psi \right \rangle \left \langle \tilde{\Psi}|\Psi \right
\rangle )^{2}],  \tag{C13}
\end{align}%
where

\begin{equation}
\left \langle \tilde{\Psi}|\tilde{\Psi}\right \rangle =N_{3}^{2}\frac{%
\partial ^{2m}}{\partial t^{m}\partial s^{m}}\left( X_{2}X_{3}-X_{4}\right)
X_{5}|_{s=t=0},  \tag{C14}
\end{equation}%
\begin{equation}
\left \langle \tilde{\Psi}|\Psi \right \rangle =N_{3}^{2}\frac{\partial ^{2m}%
}{\partial t^{m}\partial s^{m}}X_{2}X_{5}|_{s=t=0},  \tag{C15}
\end{equation}%
\begin{equation}
\left \langle \Psi |\tilde{\Psi}\right \rangle =N_{3}^{2}\frac{\partial ^{2m}%
}{\partial t^{m}\partial s^{m}}X_{3}X_{5}|_{s=t=0},  \tag{C16}
\end{equation}%
\begin{align}
\left \langle \Psi \right \vert n\left \vert \Psi \right \rangle &
=N_{3}^{2}\sinh ^{2}g  \notag \\
& \times \frac{\partial ^{2m}}{\partial t^{m}\partial s^{m}}\left(
X_{6}+1\right) X_{5}|_{s=t=0},  \tag{C17}
\end{align}%
\begin{align}
\left \langle \Psi \right \vert \Delta n_{a}^{2}\left \vert \Psi \right
\rangle & =N_{3}^{2}\sinh ^{4}g  \notag \\
& \times \frac{\partial ^{2m}}{\partial t^{m}\partial s^{m}}\left(
X_{6}^{2}+4X_{6}+2\right) X_{5}|_{s=t=0}  \notag \\
& +N_{3}^{2}\sinh ^{2}g  \notag \\
& \times \frac{\partial ^{2m}}{\partial t^{m}\partial s^{m}}\left(
X_{6}+1\right) X_{5}|_{s=t=0}  \notag \\
& -N_{3}^{4}\sinh ^{4}g  \notag \\
& \times \lbrack \frac{\partial ^{2m}}{\partial t^{m}\partial s^{m}}\left(
X_{6}+1\right) X_{5}|_{s=t=0}]^{2},  \tag{C18}
\end{align}%
\begin{align}
\left \langle \tilde{\Psi}\right \vert n\left \vert \Psi \right \rangle &
=N_{3}^{2}\sinh ^{2}g  \notag \\
& \times \frac{\partial ^{2m}}{\partial t^{m}\partial s^{m}}X_{3}\left(
X_{6}+2\right) X_{5}|_{s=t=0},  \tag{C19}
\end{align}%
\begin{align}
\left \langle \Psi \right \vert n\left \vert \tilde{\Psi}\right \rangle &
=N_{3}^{2}\sinh ^{2}g  \notag \\
& \times \frac{\partial ^{2m}}{\partial t^{m}\partial s^{m}}X_{2}\left(
X_{6}+2\right) X_{5}|_{s=t=0},  \tag{C20}
\end{align}%
and%
\begin{equation}
N_{3}=\left( \frac{\partial ^{2m}}{\partial t^{m}\partial s^{m}}%
X_{5}|_{s=t=0}\right) ^{-\frac{1}{2}},  \tag{C21}
\end{equation}%
as well as%
\begin{align}
X_{1}& =\frac{1}{2}\sinh 2g\left( -e^{\left( -i\phi \right) }\sqrt{\eta }%
+1\right) ,  \tag{C22} \\
X_{2}& =\left( X_{1}^{\ast }-\frac{1}{2}\sinh 2g\right) \left( is\beta
+istX_{1}\right) ,  \tag{C23} \\
X_{3}& =\left( \frac{1}{2}\sinh 2g-X_{1}\right) \left( it\beta
+itsX_{1}^{\ast }\right) ,  \tag{C24} \\
X_{4}& =st\left( \frac{1}{2}\sinh 2g-X_{1}\right)  \notag \\
& \times \left( X_{1}^{\ast }-\frac{1}{2}\sinh 2g\right) ,  \tag{C25} \\
X_{5}& =\exp \left[ \beta \left( sX_{1}^{\ast }+tX_{1}\right) +st\left \vert
X_{1}\right \vert ^{2}\right] ,  \tag{C26} \\
X_{6}& =\left( \beta +tX_{1}\right) \left( \beta +sX_{1}^{\ast }\right) .
\tag{C27}
\end{align}

\textbf{APPENDIX\ D : INTERNAL MEAN PHOTON NUMBER}

The expression for the average number of internal photons in the equivalent
model is given by

\begin{align}
N_{T}& =\left. _{int}\left \langle \psi \right \vert \left( a^{\dagger
}a+b^{\dagger }b\right) \left \vert \psi \right \rangle _{int}\right.  \notag
\\
& =N_{4}^{2}\{_{in}\left \langle \psi \right \vert [S_{1}^{\dagger
}B_{1}^{\dagger }U_{\phi }^{\dagger }S_{2}^{\dagger }a^{\dagger
m}S_{2}a^{\dagger }  \notag \\
& aS_{2}^{\dagger }a^{m}S_{2}U_{\phi }B_{1}S_{1}]\left \vert \psi \right
\rangle _{in}\}  \notag \\
& +N_{4}^{2}\{_{in}\left \langle \psi \right \vert [S_{1}^{\dagger
}B_{1}^{\dagger }U_{\phi }^{\dagger }S_{2}^{\dagger }a^{\dagger
m}S_{2}b^{\dagger }  \notag \\
& bS_{2}^{\dagger }a^{m}S_{2}U_{\phi }B_{1}S_{1}]\left \vert \psi \right
\rangle _{in}\}  \notag \\
& =\left( G_{m}e^{n_{1}}\right) ^{-1}\times H_{m}e^{n_{2}},  \tag{D1}
\end{align}%
where%
\begin{equation}
\left \vert \psi \right \rangle _{int}=N_{4}S_{2}^{\dagger
}a^{m}S_{2}U_{\phi }B_{1}S_{1}\left \vert \psi \right \rangle _{in}\left
\vert 0\right \rangle _{a_{v_{1}}},  \tag{D2}
\end{equation}%
and

\begin{equation}
n_{1}=st\left \vert v_{1}\right \vert ^{2}+\left( sv_{1}^{\ast
}+tv_{1}\right) \beta ,  \tag{D3}
\end{equation}%
\begin{align}
n_{2}& =\left( c\cosh g+sv_{1}^{\ast }+tv_{1}+d\cosh g\right) \beta  \notag
\\
& +\left( dv_{2}^{\ast }+sv_{1}^{\ast }+tv_{1}+cv_{2}\right) \beta  \notag \\
& +\left( tv_{1}+cv_{2}\right) \times \left( dv_{2}^{\ast }+sv_{1}^{\ast
}\right)  \notag \\
& +sdv_{1}^{\ast }\cosh g+cd\sinh ^{2}g  \notag \\
& +tv_{1}\left( c\cosh g+sv_{1}^{\ast }\right) ,  \tag{D4}
\end{align}%
as well as%
\begin{equation}
v_{1}=\frac{1}{2}\sinh 2g(1-\sqrt{T}e^{-i\phi }),  \tag{D5}
\end{equation}%
\begin{equation}
v_{2}=-\sqrt{T}\sinh ge^{-i\phi },  \tag{D6}
\end{equation}%
with%
\begin{equation}
N_{4}=\left( G_{m}e^{n_{1}}\right) ^{-\frac{1}{2}}.  \tag{D7}
\end{equation}


\begin{thebibliography}{99}
\bibitem{1} V. Giovannetti, S. Lloyd, and L. Maccone, Quantum metrology,\
Phys. Rev. Lett. \textbf{96}, 010401 (2006).

\bibitem{2} F. Hudelist, J. Kong, C. J. Liu, J. T. Jing, Z. Y. Ou, and W. P.
Zhang, Quantum metrology with parametric amplifier-based photon correlation
interferometers,\ Nat. Commun. \textbf{5}(1), 3049 (2014).

\bibitem{3} B. M. Escher, R. L. de Matos Filho, and L. Davidovich, General
framework for estimating the ultimate precision limit in noisy
quantum-enhanced metrology,\ Nat. Phys. \textbf{7}(5), 406--411 (2011).

\bibitem{4} J. J. Bollinger, W. M. Itano, D. J. Wineland, and D. J. Heinzen,
Optimal frequency measurements with maximally correlated states,\ Phys. Rev.
A \textbf{54}, R4649 (1996).

\bibitem{5} X. Zuo, Z. Yan, Y. Feng, J. Ma, X. Jia, C. Xie, and K. Peng,
Quantum interferometer combining squeezing and parametric amplification,\
Phys. Rev. Lett. \textbf{124}(17), 173602 (2020).

\bibitem{6} R. Birrittella, J. Mimih, and C. C. Gerry, Multiphoton quantum
interference at a beam splitter and the approach to Heisenberg-limited
interferometry,\ Phys. Rev. A \textbf{86}(6), 063828 (2012).

\bibitem{7} C. M. Caves, Quantum-mechanical noise in an interferometer,\
Phys. Rev. D \textbf{23}, 1693 (1981).

\bibitem{8} P. M. Anisimov, G. M. Raterman, A. Chiruvelli, W. N. Plick, S.
D. Huver, H. Lee, and J. P. Dowling, Quantum metrology with two-mode
squeezed vacuum: parity detection beats the heisenberg limit,\ Phys. Rev.
Lett. \textbf{104}, 103602 (2010).

\bibitem{b1} R. Demkowicz-Dobrza\'{n}ski, K. Banaszek, and R. Schnabel,
Fundamental quantum interferometry bound for the squeezed-light-enhanced
gravitational wave detector GEO 600,\ Phys. Rev. A \textbf{88}(4), 041802
(2013).

\bibitem{b2} J. Aasi, J. Abadie, B. P. Abbott, R. Abbott, T. D. Abbott, M.
R. Abernathy, C. Adams, T. Adams, P. Addesso, R. X. Adhikari $et$ $al.$,
Enhanced sensitivity of the LIGO gravitational wave detector by using
squeezed states of light, Nat. Photonics \textbf{7}, 613 (2013).

\bibitem{b3} M. Tse, H. Yu, N. Kijbunchoo, A. Fernandez-Galiana, P. Dupej,
L. Barsotti, C. D. Blair, D. D. Brown, S. E. Dwyer, A. Effler $et$ $al.$,
Quantum-enhanced advanced LIGO detectors in the era of gravitational-wave
astronomy, Phys. Rev. Lett. \textbf{123}, 231107 (2019).

\bibitem{m1} C. H. Oh, S. S. Zhou, Y. Wong, and L. Jiang, Quantum limits of
superresolution in a noisy environment, Phys. Rev. Lett. \textbf{126}(12),
120502 (2021).

\bibitem{m2} M. Tsang, R. Nair, and X. M. Lu, Quantum theory of
superresolution for two incoherent optical point sources, Phys. Rev. X
\textbf{6}(3), 031033 (2016).

\bibitem{m3} R. Nair and M. Tsang, Interferometric superlocalization of two
incoherent optical point sources, Opt. Express \textbf{24}(4), 3684--3701
(2016).

\bibitem{g1} M. A. Taylor, J. Janousek, V. Daria, J. Knittel, B. Hage, H.-A.
Bachor, and W. P. Bowen, Biological measurement beyond the quantum limit,
Nat. Photonics \textbf{7}, 229 (2013).

\bibitem{g2} M. A. Taylor and W. P. Bowen, Quantum metrology and its
application in biology, Phys. Rep. \textbf{615}, 1 (2016).

\bibitem{9} S. L. Braunstein and P. V. Loock, Quantum information with
continuous variables,\ Rev. Mod. Phys. \textbf{77}(2), 513--577 (2005).

\bibitem{10} V. Giovannetti, S. Lloyd, and L. Maccone, Quantum-enhanced
measurements: Beating the standard quantum limit,\ Science \textbf{306}%
(5700), 1330--1336 (2004).

\bibitem{11} V. Giovannetti, S. Lloyd, and L. Maccone, Advances in quantum
metrology,\ Nat. Photonics \textbf{5}(4), 222--229 (2011).

\bibitem{12} Z. Y. Ou and X. Li, Quantum SU(1,1) interferometers: Basic
principles and applications,\ APL Photonics \textbf{5}(8), 080902 (2020).

\bibitem{13} Y. K. Xu, S. K. Chang, C. J. Liu, L. Y. Hu, and S. Q. Liu,
Phase estimation of an SU(1,1) interferometer with a coherent superposition
squeezed vacuum in a realistic case,\ Opt. Express \textbf{30}(21),
38178(2022).

\bibitem{14} D. Li, B. T. Gard, Y. Gao, C. H. Yuan, W. P. Zhang, H. Lee, and
J. P. Dowling, Phase sensitivity at the Heisenberg limit in an SU(1,1)
interferometer via parity detection,\ Phys. Rev. A \textbf{94}(6), 063840
(2016).

\bibitem{16} C. W. Helstrom, Quantum detection and estimation theory,
(Academic, 1976), Vol. 123.

\bibitem{17} M. J. Holland and K. Burnett, Interferometric detection of
optical phase shifts at the heisenberg limit,\ Phys. Rev. Lett. \textbf{71},
1355 (1993).

\bibitem{19} Z. Y. Ou, Complementarity and fundamental limit in precision
phase measurement,\ Phys. Rev. Lett. \textbf{77}, 2352 (1996).

\bibitem{20} J. Joo, W. J. Munro, and T. P. Spiller, Quantum metrology with
entangled coherent states,\ Phys. Rev. Lett. \textbf{107}, 083601 (2011).

\bibitem{21} Y. Israel, S. Rosen, and Y. Silberberg, Supersensitive
polarization microscopy using NOON states of light,\ Phys. Rev. Lett.
\textbf{112}, 103604 (2014).

\bibitem{22} R. A. Campos, C. C. Gerry, and A. Benmoussa, Optical
interferometry at the Heisenberg limit with twin Fock states and parity
measurements,\ Phys. Rev. A \textbf{68}, 023810 (2003).

\bibitem{23} T. Ono and H. F. Hofmann, Effects of photon losses on phase
estimation near the Heisenberg limit using coherent light and squeezed
vacuum,\ Phys. Rev. A \textbf{81}, 033819 (2010).

\bibitem{24} Z. K. Zhao, H. Zhang, Y. B. Huang, and L. Y. Hu, Phase
estimation of a Mach-Zehnder interferometer via the Laguerre excitation
squeezed state, Opt. Express \textbf{31}, 17645 (2023).

\bibitem{25} H. Zhang, W. Ye, C. P. Wei, C. J. Liu, Z. Y. Liao, and L. Y.
Hu, Improving phase estimation using number-conserving operations,\ Phys.
Rev. A \textbf{103}(5), 052602 (2021).

\bibitem{26} Y. K. Xu, T. Zhao, Q. Q. Kang, C. J. Liu, L. Y. Hu, and S. Q.
Liu, Phase sensitivity of an SU(1,1) interferometer in photo-loss via photon
operations,\ Opt. Express \textbf{31}(5), 8414(2023).

\bibitem{27} Q. K. Gong, X. L. Hu, D. Li, C. H. Yuan, Z. Y. Ou, and W. P.
Zhang, Intramode-correlation-enhanced phase sensitivities in an SU(1,1)
interferometer,\ Phys. Rev. A \textbf{96}(3), 033809 (2017)

\bibitem{28} S. Ataman, Optimal Mach-Zehnder phase sensitivity with Gaussian
states,\ Phys. Rev. A \textbf{100}, 063821 (2019).

\bibitem{29} B. Yurke, S. L. McCall, and J. R. Klauder, SU(2) and SU(1,1)
interferometers,\ Phys. Rev. A \textbf{33}(6), 4033--4054 (1986).

\bibitem{30} J. D. Zhang, C. L. You, C. Li, and S. Wang, Phase sensitivity
approaching the quantum Cram\'{e}r-Rao bound in a modified SU(1,1)
interferometer,\ Phys. Rev. A \textbf{103}, 032617 (2021).

\bibitem{31} K. Zhang, Y. H. Lv, Y. Guo, J. T. Jing, and W. M. Liu,
Enhancing the precision of a phase measurement through phase-sensitive
non-Gaussianity,\ Phys. Rev. A \textbf{105}, 042607 (2022).

\bibitem{32} J. Kong, Z. Y. Ou, and W. P. Zhang, Phase-measurement
sensitivity beyond the standard quantum limit in an interferometer
consisting of a parametric amplifier and a beam splitter,\ Phys. Rev. A
\textbf{87} 023825 (2013).

\bibitem{a1} O. Seth, X. F. Li, H. N. Xiong, J. Y. Luo, and Y. X. Huang,
Improving the phase sensitivity of an SU(1,1) interferometer via a nonlinear
phase encoding,\ J. Phys. B: At., Mol. Opt. Phys. \textbf{53}(20), 205503
(2020).

\bibitem{a2} S. K. Chang, W. Ye, H. Zhang, L. Y. Hu, J. H. Huang, and S. Q.
Liu, Improvement of phase sensitivity in an SU(1,1) interferometer via a
phase shift induced by a Kerr medium,\ Phys. Rev. A \textbf{105}(3), 033704
(2022).

\bibitem{a3} S. Ataman, Quantum Fisher information maximization in an
unbalanced interferometer,\ Phys. Rev. A \textbf{105}, 012604 (2022).

\bibitem{a4} S. Ataman, K. K. Mishra, Quantum Fisher information
maximization in an unbalanced lossy interferometer,\ Phys. Rev. A \textbf{109%
}, 062605 (2024).

\bibitem{a5} M. Jarzyna, R. Demkowicz-Dobrzanski, Quantum interferometry
with and without an external phase reference,\ Phys. Rev. A \textbf{85},
011801 (2012).

\bibitem{33} S. S. Liu, Y. B. Lou, J. Xin, and J. T. Jing, Quantum
enhancement of phase sensitivity for the bright-seeded SU(1,1)
interferometer with direct intensity detection,\ Phys. Rev. Appl. \textbf{10}%
(6), 064046 (2018).

\bibitem{34} M. Manceau, G. Leuchs, F. Khalili, and M. Chekhova, Detection
loss tolerant supersensitive phase measurement with an SU(1,1)
interferometer,\ Phys. Rev. Lett. \textbf{119}(22), 223604 (2017).

\bibitem{35} W. Du, J. Jia, J. F. Chen, Z. Y. Ou, and W. P. Zhang, Absolute
sensitivity of phase measurement in an SU(1,1) type interferometer,\ Opt.
Lett. \textbf{43}(5), 1051 (2018).

\bibitem{36} S. S. Szigeti, R. J. Lewis-Swan, and S. A. Haine, Pumped-up
SU(1,1) interferometry, Phys. Rev. Lett. \textbf{118}, 150401 (2017).

\bibitem{a7} D. Li, C. H. Yuan, Z. Y. Ou, and W. P. Zhang, The phase
sensitivity of an SU(1,1) interferometer with coherent and squeezed-vacuum
light,\ New J. Phys. \textbf{16}(7), 073020 (2014).

\bibitem{37} N. Namekata, Y. Takahashi, G. Fujii, D. Fukuda, S. Kurimura,
and S. Inoue, Non-Gaussian operation based on photon subtraction using a
photon-number-resolving detector at a telecommunications wavelength,\ Nat.
Photon. \textbf{4}, 655 (2010).

\bibitem{38} G. S. Agarwal and K. Tara, Nonclassical properties of states
generated by the excitations on a coherent state,\ Phys. Rev. A \textbf{43},
492 (1991).

\bibitem{39} L. Y. Hu and Z. M. Zhang, Statistical properties of coherent
photon-added two-mode squeezed vacuum and its inseparability,\ J. Opt. Soc.
Am. B \textbf{30}, 518 (2013).

\bibitem{40} A. Zavatta, V. Parigi, and M. Bellini, Experimental
nonclassicality of single-photon-added thermal light states,\ Phys. Rev. A
\textbf{75}, 052106 (2007).

\bibitem{41} Y. Ouyang, S. Wang, and L. J. Zhang, Quantum optical
interferometry via the photon-added two-mode squeezed vacuum states,\ J.
Opt. Soc. Am. B \textbf{33}, 1373 (2016).

\bibitem{42} J. Xin, Phase sensitivity enhancement for the SU(1,1)
interferometer using photon level operations,\ Opt. Express \textbf{29}(26),
43970(2021).

\bibitem{44} M. Dakna, L. Kn\"{o}ll, and D.-G. Welsch, Photon-added state
preparation via conditional measurement on a beam splitter,\ Opt. Commun.
\textbf{145}(1-6), 309--321 (1998).

\bibitem{45} L. L. Guo, Y. F. Yu, and Z. M. Zhang, Improving the phase
sensitivity of an SU(1,1) interferometer with photon-added squeezed vacuum
light,\ Opt. Express \textbf{26}(22), 29099 (2018).

\bibitem{46} X. Y. Hu, C. P. Wei, Y. F. Yu, and Z. M. Zhang, Enhanced phase
sensitivity of an SU(1,1) interferometer with displaced squeezed vacuum
light,\ Front. Phys. \textbf{11}, 114203 (2016).
\end{thebibliography}
\end{document}